Graphical Table of Contents

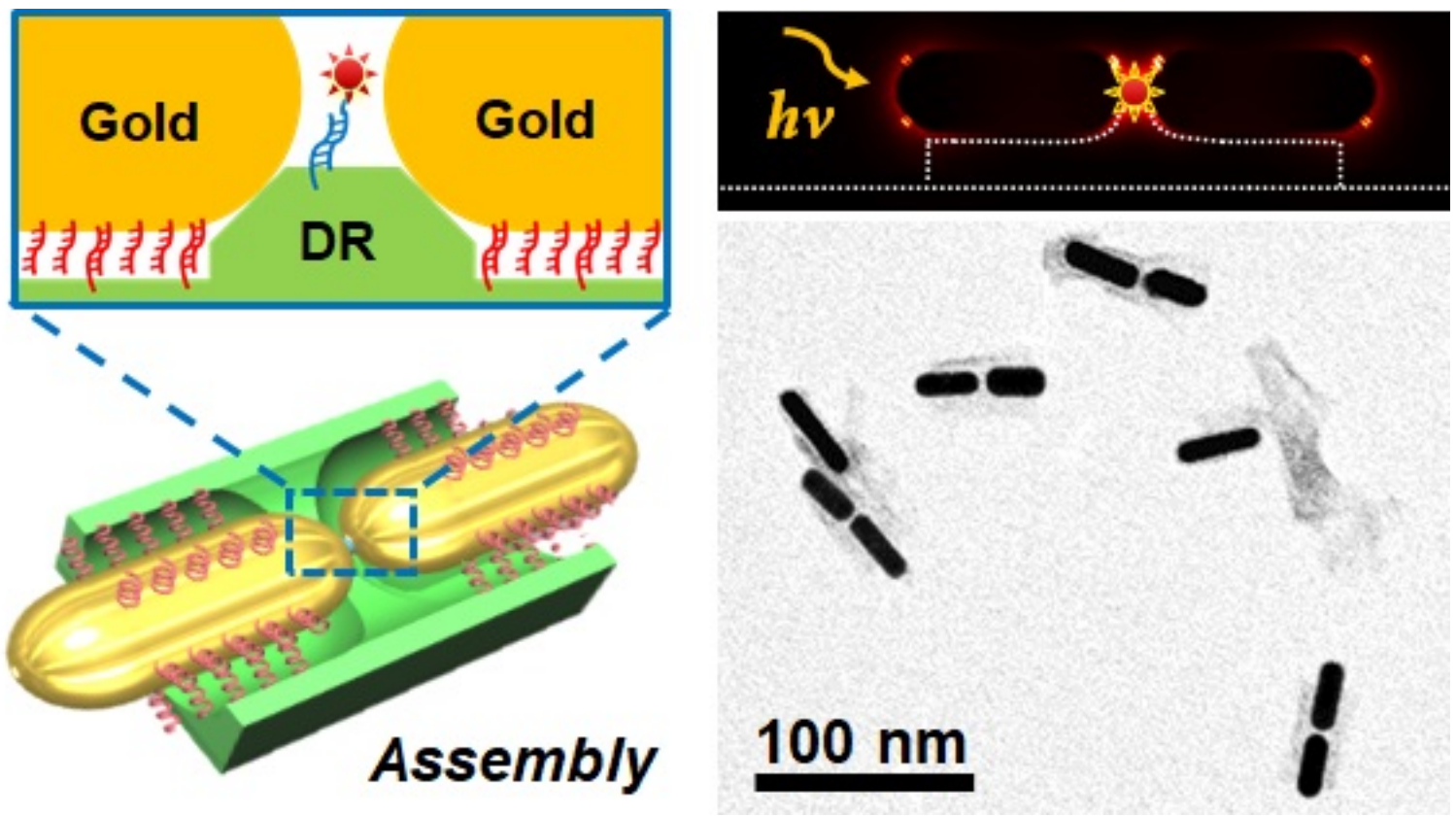

A novel strategy to deterministically place single emitters in sub-5 nanometer optical nanocavities formed by gold nanorod dimers using three-dimensional DNA origami. Such a system is highly desirable in studying light-matter interactions.

# Deterministic Assembly of Single Emitters in Sub-5 Nanometer Optical Cavity Formed by Gold Nanorod Dimers on Three-Dimensional DNA Origami

*Zhi Zhao,*[†,§,⊥] *Xiahui Chen,*[§] *Jiawei Zuo,*[§] *Ali Basiri,*[§] *Shinhyuk Choi,*[§] *Yu Yao,*[§] *Yan Liu,*[*,†,‡] *and Chao Wang*[*,†,§]

[†]Center for Molecular Design and Biomimetics at the Biodesign Institute, Arizona State University, Tempe, Arizona 85287, United States

[‡]School of Molecular Sciences, Arizona State University, Tempe, AZ 85287, United States.

[§]School of Electrical, Computer and Energy Engineering, Arizona State University, Tempe, AZ 85287, United States

[⊥]Faculty of Materials and Manufacturing, Key Laboratory of Advanced Functional Materials, Education Ministry of China, Beijing University of Technology, Beijing 100124

*E-mail (Y. Liu): Yan_Liu@asu.edu.

*E-mail (C. Wang): wangch@asu.edu.

## ABSTRACT

Controllable strong interactions between a nanocavity and a single emitter is important to manipulating optical emission in a nanophotonic systems but challenging to achieve. Here a three-dimensional DNA origami, named as DNA rack (DR) is proposed and demonstrated to deterministically and precisely assemble single emitters within ultra-small plasmonic nanocavities formed by closely coupled gold nanorods (AuNRs). The DR uniquely possesses a saddle shape with two tubular grooves that geometrically allows a snug fit and linearly align two AuNRs with a bending angle <10°. It also includes a spacer at the saddle point to maintain the gap between AuNRs as small as 2-3 nm, forming a nanocavity estimated to be 20 nm$^3$ and an experimentally measured $Q$ factor of 7.3. A DNA docking strand is designed at the spacer to position a single fluorescent emitter at nanometer accuracy within the cavity. Using Cy5 as a model emitter, a ~30-fold fluorescence enhancement and a significantly reduced emission lifetime (from 1.6 ns to 670 ps) were experimentally verified, confirming significant emitter-cavity interactions. This DR-templated assembly method is capable of fitting AuNRs of variable length-to-width aspect ratios to form anisotropic nanocavities and deterministically incorporating different single emitters, thus enabling flexible design of both cavity resonance and emission wavelengths to tailor light-matter interactions at nanometer scale.

## INTRODUCTION

The ability of nanophotonic structures to tailor and control light emission is highly attractive in enhanced imaging, data encryption, and ultra-compact circuitry.[1-3] One fundamental question is to control and strengthen the interaction between single quanta of light (photons emitted from an emitter) and single entities of matter (optical cavities) to engineer the optical coupling.[4, 5] For example, room-temperature (RT) quantum optics has been demonstrated by a number of methods[5-7] following an optical cavity design principle to maximize the "figure of merit" ($Q/\sqrt{V}$),[8] where $V$ is the mode volume and $Q$ is the cavity quality factor. A small $V$ is favorable for enhancing the field-intensity and the coupling strength ($g$) of emitter in the cavity, while a large $Q$ indicates a minimized energy loss. Given an emitter with an oscillator strength $f$, a maximal $Q/\sqrt{V}$ would lead to optimal coupling strength, as $g \propto f \cdot Q/\sqrt{V}$.

Traditionally, micro- or nano-structured dielectric cavities have been utilized for their high $Q$ factors, such as micro-pillars,[4] microdisks[9] and photonic crystals.[6,10,11] Recently, plasmonic nanocavities, such as nanowires,[12] nanodisk dimers[7] and nanoparticles (cubes or spheres) coupled to a mirror,[13,14] have shown significant enhancement of coupling strength as a results of greatly reduced mode volume.[13,14] However, it still remains challenging to achieve deterministic placement of single emitters in plasmonic or dielectric cavities in a reproducible manner.

To date, a number of methods had been proposed to align the emitter-cavity systems, including tip-based mechanical movement, aligning cavity on tracked emitters, etc.[15,16] Nevertheless, these methods relied on mechanical or lithographic alignments, making it impractical to achieve reproducible and accurate placement of an individual single emitter molecule within a nanometer-scale cavity. Additionally, the spatial alignment accuracy, even using advanced electron-beam system, was practically restricted to tens of nanometers. Over such a length scale, unfortunately, the electromagnetic wave distribution in nanophotonic cavities could change drastically, thus limiting fundamental studies and device applications.

DNA origami (DO)[17,18], unlike conventional nanopatterning methods, enables bottom-up assembly to integrate organic emitters and plasmonic structures with nanometer accuracy at one step.[19,20] During the assembly process, hundreds of short "staple" DNA strands can be programmed to fold a long scaffold strand into a relatively rigid structural template with a set of surface anchor strands at selective positions to specifically attach single stranded DNA-modified plasmonic nanoparticles and quantum emitters. The bottom-up self-assembly process eliminates complex and costly top-down alignment steps. Additionally, such a strategy allows emitters to be deterministically positioned in the cavity at nanometer precision[17] (e.g. 6 nm pixel size) that is well beyond the conventional lithographical capability.[21,22] In one example, a nanocavity of a mode volume of ~ 200 $nm^3$ was achieved using surface coupled nanospheres.[23] In another demonstration, a gap of >5 nm was formed between two gold nanospheres on a planar DO.[24] However, these demonstrations still had difficulties to achieve significant field enhancement and optical coupling at the single emitter level,[23,24] which was attributed to the design challenges on achieving small-mode-volume optical cavity and precisely placing a single emitter on the same DO template at the center of the nanocavity.

Here we demonstrate deterministic placement of a single fluorescent emitter into the nanogap between a well-aligned gold nanorods (AuNRs) dimer assembled on a three-dimensional (3D) DO. This nanodevice displays highly enhanced emitter-cavity interactions due to improved optical coupling. The pair of AuNRs is aligned tip-to-tip at a close distance of about 2-3 nm via specifically designed saddle-shaped DO template featured with two tubular grooves, allowing geometrical snug fit to assemble two AuNRs with a specific width matching the inner dimension of the grooves. Such a design can greatly reduce mode volume $V$, improve $Q$ factor, and accordingly boost the coupling strength $g$.[25,26] Importantly, this DO template can accommodate AuNRs of the same diameter but varying length, and thus makes it possible to create nanocavities between a pair of AuNRs of different aspect ratios (AR).[27]

In this way, the cavity resonance wavelength is tuned to enable the systematic analysis of cavity-emitter interaction. Additionally, one single emitter labeled with single strand DNA (ssDNA) is incorporated precisely at the center of the AuNR dimer by hybridization with a programmed docking strand extending from the saddle point of the DO template. The positioning of the single emitter in the center of the nanocavity is expected to be within single digit nm precision. By incorporating Cy5 as the emitter, and using AuNR dimer of AR=2.8 and width =12 nm (effective mode volume ~20 $nm^3$), we demonstrated deterministic self-assembled coupling between plasmonic nanocavity and emitter with a high figure of merit $Q/\sqrt{V}$ (measured to be 1.6 $nm^{-3/2}$). We observed experimentally enhanced fluorescence emission (>30 folds) and significantly reduced emission lifetime (from 1.6 ns down to 670 ps), compared with that of free Cy5 in solution.

## RESULTS AND DISCUSSION

**3D DNA origami template for assembly of AuNR nanocavities.** Plasmonic nanocavities assembled from nanoparticles had been used for studying the cavity-emitter interactions.[13,14,23,24] Maximizing the coupling strength between an emitter and the plasmonic nanocavity requires a few design considerations. First, in order to overcome the inevitable optical loss and reach the strong emitter-cavity coupling,[13] the effective optical mode volume $V_{eff}$ of the plasmonic nanocavity needs to be minimized, *e.g.* to as small as $10^{-6}$. Here $V_{eff}$ is defined as $V_{eff} = V/V_{\lambda} = V(\frac{n}{\lambda})^3$, where $\lambda$ is the emission wavelength, $n$ is the medium refractive index, and $V$ is the cavity optical mode volume. The minimization of $V_{eff}$ can be achieved by reducing $V$ to create an ultrasmall gap between the metallic nanoparticles as small as $\lambda/100$, *e.g.* 5 nm.[13] However, such a small gap size is challenging to achieve using conventional design and fabrication strategies (Table S1). In particular, it is feasible to separate plasmonic nanoparticle dimers, or place nanoparticles on a metal mirror, with such a small distance *via* inserting a DO in between (Figure S1), but the as-obtained gap sizes are constrained by the intrinsic dimensions of the double strand DNA (dsDNA) and the length of the linker molecules, thus it is challenging to achieve nanocavities with ultra-small (e.g. <5 nm) dimensions.[23,24] Moreover, most of the conventional methods cannot guarantee accurate and deterministic placement of the emitter to the nanocavity hotspot, which is essential for maximizing the emitter-cavity interaction.

Here, we address these challenges by employing a novel 3D DNA origami design, named as DNA rack (DR), to guide the assembly of single emitter precisely into the nanocavity formed by AuNRs with distances as small as 3 nm. The single emitter can be accurately and deterministically positioned at the nanocavity hotspot to maximize the emitter-cavity interaction. Moreover, our method allows tuning the plasmonic resonance of the nanocavities to match the emitter wavelength with minimal redesign efforts, which is also essential for strong enhancement of optical emission at a reduced $V_{eff}$.[28,29]

The DNA rack is a 3D DO template featuring two tubular grooves separated by a saddle-shaped spacer (Figure 1A), which allows a geometrical snug fit of a pair of AuNRs of specific width, one in each groove, in order to both confine the orientation of AuNRs and define the gap size (Figure 1B). In particular, AuNRs in a dimer were forced ~2 nm apart (tip to tip) by the steric effect of the DR geometry and specially designed spatial distribution of AuNR docking strands inside the grooves (Figure S2). The deterministic placement of a single emitter at the center of the gap was achieved by docking one emitter-conjugated DNA strand to the complementary strand (strand 'Docking' in Table S1) at the top-center position of the spacer by DNA hybridization. Here the 'Docking' stand was carefully designed, whose 3' end protruded vertically to DR (Figure 1A) and form a 10-base-pair hybridization with the emitter-conjugated DNA strand. No single-stranded region was left in 'Docking'. Thus we expected that the displacement of the emitter should be mainly from the dangling of the single-bond linker between the emitter and DNA. In our case this moiety was short than 1 nm, which ensured the high placement accuracy of emitters.

The assembly of anisotropic plasmonic nanocavities using an AuNR dimer is more desirable compared with using dimers of gold nanosphere (AuNS) of the same volume and gap size. This is due to the following reasons as revealed by our finite-difference time-domain (FDTD) simulations (Figure S3).

First, AuNS dimers (diameter 18 nm, gap 4 nm) provide only moderate field-enhancement factor $EF = |E/E_0|^2$, where $E$ and $E_0$ are the intensity of electric field inside the nanocavity and in the incident light, respectively (Figure S3A-S3B). In comparison, the value of $EF$ could reach an order of magnitude higher for a nanocavity formed by two coupled AuNRs (diameter 12 nm, length 30 nm, gap 4 nm) (Figure S3D-S3E).

Second, the plasmonic nanocavity formed by end-to-end coupled AuNR dimer is more advantageous in reducing non-radiative loss and enhancing light-matter interactions compared with AuNS dimers. Our simulation results (Figure S3E & S3B) showed the AuNR dimer possessed an ultra-small mode volume ($V$ = 20 nm$^3$, see method section for more detailed information), *i.e.* about 10 times smaller than that of AuNS dimers (200 nm$^3$), and an approximately twice as high $Q$ factor in the calculation (14.6 for AuNR and 8.0 for AuNS).

Additionally, anisotropic nanocavity design also features greater flexibility to achieve a broad optical resonance tuning by simply changing the nanorod geometry, in particular its AR (Figure S3F).[30] This allows programmable resonance detuning using one DR design, which is important to a systematic study of the coupled emitter-cavity system. In theory, the cavity resonance of coupled AuNR dimers could be tuned from 580 nm to over 780 nm by using AuNRs of a fixed width of 12 nm but varied lengths from 20 to 50 nm (Figure S3F). The actual resonance tuning accuracy may be limited by the deviation of AuNRs' size and assembly geometry, yet should be fairly enough for emitters with well separated emissions (e.g. tens of nm). This covers a series of widely used organic dyes, including Cy3, Cy5, Cy7, *etc*.[31] In comparison, nanocavities formed by dimers of AuNSs with 20 nm to 50 nm diameter displayed only minimal resonance change from 530 to 540 nm (Figure S3C), and thus their use could be limited by the availability of emitters for efficient coupling.

Furthermore, our simulation indicated that anti-crossing behavior with a hypothetical emitter (emission maximum at 773 nm, similar as Cy7) could only be observed in AuNR based anisotropic nanocavities (white arrow in Figure S4A), but not in AuNS based nanocavities (Figure S4B) using otherwise identical conditions.

Lastly, from the fabrication point of view, the same DR design could be directly used to assemble AuNRs of the same width but varying lengths (see next section for more discussions), which greatly simplified both the design and experiments. In contrast, substantial modifications in the design would be necessary in order to assemble AuNSs of different diameters by using the same strategy to control a small inter-particle distance.

**Assembly of AuNR nanocavities.** The DR was designed with Cadnano 2.0 (Figure S5) and prepared following previous established annealing protocol.[32] Single stranded, circular viral DNA from M13mp18 was utilized as the scaffold, which was folded by a total number of 217 staple strands into the designed 3D shape with well-defined dimensions. After annealing, the product was purified with a 100 kD Amicon micro-spin filter under centrifugation at 8000 rpm for 10 min to remove the excess staple strands. Native agarose gel electrophoresis showed that the purified sample was a pure product, displaying a single migration band between 4k and 5k bp markers (Figure 1F). The purified DR sample was further examined under transmission electron microscope (TEM), showing a high yield (~90%) and a morphology well matched to our design (Figure 1G). For example, the measured length and width of DR was 57.5±1.4 nm and 29.7±0.9 nm, while the expected length and width from the design was 58.3 nm and 30.0 nm, respectively. The detailed features of the DR, *e.g.* the grooves and the central spacer, were visible in magnified TEM images (Figure 1G, inset) and atomic force microscope (AFM) images (Figure 1H). The width of the groove was measured to be ~23.1± 0.8 nm from TEM, slightly smaller than expected (25 nm), which could be attributed to possible tilting or curving of the DR.

To successfully assemble the DR-guided AuNR cavity with an optimal yield, the critical dimensions of the DR template and the AuNRs were designed to match. Here we followed a previously developed synthetic strategy to prepare AuNRs of designed lengths and widths so they could fit the DR.[33] Briefly, a silver seed solution was added to a mixture containing $AgNO_3$ and $HAuCl_4$ for a controlled seeded-growth of AuNRs. By adjusting the concentrations of $HAuCl_4$ and $AgNO_3$, the width and length of the AuNRs could be tuned individually (Figure S6). The successful preparation of a series of AuNRs of different ARs was confirmed by TEM imaging (Figure 2A-2D & S7), showing a relative standard deviation (SD%) of 8-9% in both length and width. Clearly, the tuning of the AuNR length from 30 nm (Figure 2A & 2C) to 34 nm (Figure 2B & 2D) did not affect the NR width and thus the assembly to DNA origami. AuNRs of 12 nm width were chosen to construct the plasmonic nanocavities, whose geometry best fits the inner dimensions of the grooves of the DR.

To prepare the AuNRs for assembly with the DR, the AuNRs of the selected dimensions were surface-modified with a layer of 12-mer poly(T) capping strands, by incubating the AuNRs with thiol-terminated DNA capping strands and aging at elevated salt concentrations.[34] UV-Vis spectra of the DNA modified AuNRs showed a small red shift with regard to the unmodified samples (Figure S8), demonstrating the successful surface modification. NuPACK calculations indicated that such a capping strand could maintain a tight binding to its complementary strand below 45 °C, which is important to the formation and stability at RT. Although each nucleotide in a ssDNA was estimated to be ~ 0.56 nm long,[35] the short persistent length of ssDNA (~ 2 nm) should make the capping layer flexible[36,37] enough for the DNA-capped AuNRs to fit into the grooves on the DR. Upon hybridization to form dsDNA on the DR, the docking strands can be intuitively understood as rigid rods to bind AuNRs to the DR, considering a large persistent length of dsDNA of ~ 50 nm.[38] Such a rigidity is important to control of the DNA origami dimensions at nanometer scales. In our design, the longitudinal axis of AuNRs is estimated to be ~3 nm above the top of the spacer, so that the emitter is expected to be located at the center of the nanocavity gap, where the highest EF is expected (Figure 1B).

The DNA capped AuNR (*e.g.* AR=2.8) was mixed with the purified DR, and subjected to multiple annealing treatments to assemble the plasmonic nanocavities. Fluorescence melting measurement (Figure S9) indicated that the melting temperature of the AuNR docking strands hybridized with the complementary capping strands was around 42 °C, much lower than that of the DR (~62 °C). Therefore, the DR structure should remain intact during annealing. Here we set the starting annealing temperature at 45 °C, and slowly lowered it down to 30 °C in ~15 h. In order to facilitate the binding of AuNRs and the formation of AuNR linear dimers, the annealing cycle was repeated 4 times to obtain the most energetically stable product.

After annealing, the product was characterized with a microscope-coupled UV-Vis spectrometer (Figure S10). The working principle of each type of measurements is detailed in Supplementary section 5. The major absorption bands of the AuNR monomer peaked at 520 nm and 640 nm, which are the transversal and longitudinal plasmonic resonance modes, respectively (Figure 2E), *i.e.* polarization along the short- and long-axis of the AuNRs. The AuNRs dimer product showed an additional band at 712 nm (the blue peak in Figure 2E), representing the optical coupling of AuNRs from the small nanogaps and indicating the successful assembly of the AuNR dimers. Spectral deconvolution revealed a small peak at 640 nm (the green peak in Figure 2E) from the assembly products, indicating a small percentage of remaining AuNR monomers. TEM images also revealed that AuNR dimers formed successfully (Figure 2F & S11).

The narrow gaps between the dimers were clearly visible from high-magnification TEM images (Figure 2G). By analyzing the distribution of the gap sizes (AuNR tip to tip distance) from the images of hundreds of dimers (Figure 2H), we found that the majority were smaller than 5 nm, and about a half was in the 2-3 nm range that is ideal for the investigation of cavity-emitter interaction. To the best of our knowledge, such a narrow gap size between plasmonic nanoparticles is one of the smallest achieved with DO templated assembly,[39,40] especially for tip-to-tip assembly of AuNRs. Moreover, the major portion (~95%) of the dimer possessed a near linear conformation ($\theta<10°$), attributed partly to the high structural stability of the designed DR (Figure 2I). The demonstrated high AuNR assembly accuracy of <3 nm (Figure 2F-2I) is favorable for studying emission enhancement and data analysis across different sample batches.

**Surface immobilization of the AuNR dimer**. The as-prepared AuNR-DR complexes were immobilized onto fused silica substrate through electrostatic interactions (Figure 3A) for imaging and spectral characterizations. Following previously established protocols,[11] the surface of the substrate was first modified with a negatively charged molecular species by carboxyethylsilanetriol disodium salt, which then associated with DR (negatively charged) under the presence of $Mg^{2+}$ ions. Briefly, diluted AuNR-DR solution was drop-casted onto fused silica, incubated for 1 hour, rinsed with copious amount of Tris buffer, and stabilized with high-pH (8.9), $Mg^{2+}$ containing (15 mM) buffer. The as-prepared samples were examined under scanning electron microscope (SEM, Figure S12), revealing successful immobilization of AuNR-DR complexes with a surface density of AuNR dimers ~ 0.5 per 1 $\mu m^2$. A small amount of AuNR monomers were also observed, attributed to the excessive AuNRs used during assembly, consistent with the UV-Vis spectral analysis. Compared with their dimers, the AuNR monomer had a much lower surface density and would not be expected to significantly contribute to emission enhancement of the emitter or energy splitting, due to their much smaller field enhancement and resonance mismatch (explained in detail in the next section). Therefore, no extensive purification was conducted on the AuNR-DR complexes to avoid unnecessary sample loss. For control experiments, DR bound AuNR monomers were also prepared by removing the docking strands on one groove side of the DR (Figure S13). The products were purified, immobilized and examined using the same procedures.

**Microscopic and spectral characterization of optical emitters in anisotropic nanocavity**. We employed dark field scattering (DFS) microscopy to characterize the nanocavity-emitter coupling (Figure 3B & Supplementary section 6). The mechanism was schemed in Figure S14.

DFS spectra were each taken from a 62.5 µm × 62.5 µm area, which contained roughly 2000 AuNR dimers, estimated from the SEM images.

Surface immobilized samples prepared with AuNRs of AR=2.8 were first investigated as representatives. From the control sample, individual AuNRs (monomers on DR) with an emitter resulted in two resonance peaks in the DFS spectra (Figure 3C), at ~525 nm and ~610 nm, respectively, corresponding to the transversal and longitudinal resonance mode of AuNR monomer. The blue-shifted longitudinal peak in DFS with regard to that in the solution (Figure 2E) was due to the change in the refractive index (RI) of the media. On the other hand, the transverse resonance mode is mostly insensitive to the medium without significant change in resonance wavelength.[41] Different from AuNR monomers, AuNR dimers with an emitter exhibited more complex features from DFS spectra (Figure 3C). The transverse mode (525 nm peak) remained visible despite being less evident, while additional peaks between 550 nm to 750 nm can be de-convoluted into two Gaussian peaks, centered around 622 nm and 695 nm, respectively.

To better understand the near-field coupling, we simulated DFS of surface immobilized AuNR-DR complexes using FDTD (Figure 3E and 3F). The simulation results further confirmed that the experimentally observed scattering signals in 500-550 nm range (Figure 3C and 3D) could be mainly attributed to the transverse mode for both the AuNR monomers and dimers. Since the AuNR dimer samples were randomly oriented during deposition, both the transvers mode and the longitudinal mode were excited in the experimental. However, the transverse mode was not expected to strongly interact with the dye given the large wavelength mismatch (>140 nm) and poor near-field enhancement (only a few folds) (Figure S15). In contrary, the longitudinal mode of the AuNR dimer resulted in the maximum electromagnetic field intensity in the center of the nanocavity (Figure S3D) with a small optical mode volume (20 $nm^3$), and optimal resonance of the nanocavity mode at the excitation and emission wavelengths of the emitter.

To study the interaction of Cy5 emitter with AuNR nanocavities, we varied the AR of AuNRs between 2.5 and 2.8 (Figure 3G), and performed both DFS experiments and FDTD simulations (Figure 3 G-H). The DFS spectra in the range of 550-750 nm from both the experimental and the simulation data were fitted by two Gaussian peaks. As the AuNR AR increased, it was observed that the longer-wavelength peak gained intensity with regard to the shorter-wavelength peak. Meanwhile, the midpoint of the peak remains the same but the separation between the two peaks changed, thought due to the variation of resonance coupling between the cavity mode and the dye emission. This can be understood that a better match of the cavity resonance wavelength and the emission wavelength facilitates more efficient energy transfer between the emitted photons and the nanocavity, and accordingly decreases the non-radiative energy loss. Interestingly, the magnitude of energy difference between the two extracted peaks increased with a higher AuNR AR (Figure 3G), possibly attributed to strong excitation of the emitters in the cavity. Yet, future experiments to more broadly tune the AuNR AR, as well as to increase the oscillator strength of emitters, are needed for more complete understandings of the emitter-cavity interactions.[42]

These experimental observations were consistent with the FDTD simulations (Figure 3H), where an AR dependent emitter-cavity coupling strength was evidenced. It is noticed that the experimental results displayed broader emission linewidths than that of the simulation, possibly due to the collective scattering from many emitter-cavity systems (~2000 emitters under a 62.5 μm × 62.5 μm field of view) with a geometrical variation of the AuNR dimers, *e.g.* length of the AuNRs, the gap size, and derivation from perfectly straight alignment. The observed phenomenon is thought to be correlated with anti-crossing behavior (Figure S14 and S16); however, future work of single emitter analysis is needed to elucidate such an effect.

To further evaluate the impact of the plasmonic nanocavities on the emitter emission profiles, we investigated the fluorescence emission from Cy5 deterministically placed in the nanocavities using our customized microscope-coupled UV-Vis spectrometer (Figure 4A & Supplementary section 5). For consistency, the same sample area (62.5 μm × 62.5 μm field of view) was used for fluorescence measurements following the DFS experiments. The spectral data clearly evidenced that a stronger fluorescence signal was recorded from surface immobilized AuNR dimers, compared to that of AuNR monomer or only DR (*i.e.* no AuNRs) (Figure 4B). Due to the application of fluorescent filters, only signals between 670 and 720 nm were collected, leading to the cutoffs in the spectra. It is noticed that AuNRs may display interband and intraband transitions due to hot carrier excitation at high illuminating laser power densities (0.1-0.2 $MW/cm^2$);[43] however, such an effect is not expected to be dominant in our study given the relatively low power density produced by Xeon lamp (about 4 $W/cm^2$).

To quantify the overall fluorescence enhancement, the fluorescence signals between 670 nm and 720 nm were integrated (Figure 4C), showing that the emitters inside the AR=2.8 and AR=2.5 AuNR dimer nanocavities were about 30 and 2 times brighter than that on the DR template only without the AuNRs, respectively (Supplementary section 7). This can be understood from simulation (Figure 4D-4E) that the use of AuNR dimers (AR=2.8) resulted in a strong near-field enhancement at the nanocavities (dye-attachment site), reaching ~ 200 folds and ~900 folds at the absorption and emission maximum wavelengths of the dye, respectively. This relatively broad-band field enhancement overlapped well with both the emitter absorption and emission spectra, resulting in an enhanced excitation and an accelerated emission, simultaneously, and thus improved quantum efficiency.[44,45]

Further, the near-field enhancement of AR=2.5 dimers over the data collection window was found 3 times lower than that of the AR=2.8 nanocavities, showing that the AR=2.5 dimers could enhance the dye excitation but not contribute significantly to promoting emission (Figure 4E). This simulation results are consistent with the observed 90% lower fluorescence signal from AR=2.5 dimer cavities.

Using the experimentally extracted geometric parameters from the TEM images, simulations were performed to calculate the near-field enhancement of the nanocavity at the emission wavelength (670 nm) for different AuNR ARs (Figure S17), showing a maximum of $1.2\times10^3$ at AR=2.8. Further, using AR=2.8 AuNR dimer cavity as an example, the qualify factor ($Q$) of the nanocavity was calculated as 14.6 and measured as 7.3, the mode volume ($V$) was simulated as ~20 $nm^3$, and the Purcell factor was estimated $2.4\times10^6$ at dye emission (Supplementary section 8). This accordingly resulted in a large calculated $Q/\sqrt{V}$ value of 3.3 $nm^{-3/2}$ from simulation (or 1.6 $nm^{-3/2}$ using experimentally determined $Q$ factor), which is up to two order of magnitude greater than plasmonic cavities formed by randomly placed nanoparticles and about one order of magnitude better than that of formed by DO templated assembly (Table S1). These analyses show that the AuNR nanocavities are a unique design strategy for maximizing the light matter interactions.[46]

In contrast, Cy5 located close to the AuNR monomers did not display any observable fluorescence enhancement, compared with that from the dyes attached to DR only. Our simulation (Figure 4D-4E) showed that the field enhancement at the emitter site for the AuNR monomer was orders of magnitude smaller compared with that of AuNR dimers, i.e. ~75 and ~15 folds at the absorption and emission maxima, respectively. The small and narrow-band field enhancement as well as a larger resonance mismatch with regard to the emitter make the vicinity of AuNR monomer a much less effective nanocavity to overcome the fluorescence quenching due to nonradiative energy dissipation from the dye to the Au surface,[47,48] and therefore resulted in minimal overall enhancement of fluorescence signals.[49] Clearly, the presence of AuNP dimer nanocavity plays an important role in modulating the fluorescent emission process. Indeed, it has shown that Ag NPs can contribute to stabilize Cy3 or Cy5 dye molecules by promoting radiative decay channel and accordingly reducing the probability of photobleaching.[50]

To better quantify the impact of AuNR nanocavity on the optical emission, the lifetime ($\tau$) of the Cy5 dye with and without AuNRs was measured with a picosecond photon detector (PPD) device coupled to a spectrometer (Figure 4F) and then analyzed using DAS6 software (Supplementary section 9). Under the presence of excessive decay pathways from the cavity, a reduced lifetime of the emitter from the excitation state to the ground state was expected.[12] The $\tau$ of Cy5 on the DR without AuNRs was 1.6 ns, comparable to but slightly longer than that of free Cy5 in aqueous buffers (1.2 ns) (Figure S18), which was attributed to a stabilized fluorescence emission due to the attachment to DNA structures.[51,52] When placed in the nanocavity, Cy5 was found to have a reduced lifetime of ~670 ps from mathematical fitting, due to optical coupling to the AuNR cavity. The fitting quality has been confirmed by minimal fitting residues as well as small deviations between experimental and calculated numbers (Figure 4F & S19). This is consistent with the calculated large Purcell factor ($\sim 2.4\times10^6$), proving the effective modulation of optical emission. The reduced emission lifetime, attributed to cavity-enhanced emission, in turn is associated with higher photostability of dyes given possibly reduced possibly quenching or photobleaching, as demonstrated before using Cy5-labeled DNA in the presence of metallic Ag particles.[50]

## CONCLUSIONS

In summary, we demonstrated a novel strategy to construct nanometer-scale plasmonic nanocavities *via* DNA origami guided AuNR self-assembly. The 3D morphology of DR enables a snug fit of AuNR dimers of designated widths with varying lengths, forming ultra-small nanocavities with a nanogap as small as 2 nm between a pair of linearly aligned AuNRs. The programmability of the DR allows placement of a single docking strand right at the center of the spacer, which could be used to deterministically attach a single emitter of chosen wavelength by DNA hybridization. A number of critical experimental parameters, including the concentration of reactants, annealing conditions, and AuNR geometries, were examined to produce AuNR dimers of varying plasmonic resonance peaks centered from 630 nm to 680 nm.

By using a single Cy5 dye deterministically placed in the nanocavity, we demonstrated that this new AuNR-DR complex can be used to study the interactions between single emitters and ultrasmall optical nanocavities. This system enabled us to detune the AuNR dimer resonance by varying their geometry parameters and observe the impact of the emitter on the AuNR dimer scattering. The AuNR dimer cavity was observed to enhance the Cy5 fluorescence emission by ~30 times, accompanied by a ~3 folds reduction of the emission lifetime. These observations are supported by our FDTD simulations that showed significant near-field enhancement, very small optical mode volume (20 $nm^3$), a large Purcell factor (as high as $2.4\times10^6$), and a large figure of merit $Q/\sqrt{V}$ of 3.3 $nm^{-3/2}$, all comparable to reported plasmonic nanocavity designs.[13,23]

This novel 3D DR design provides flexibility in AuNR assembly and emitter selection, making it a potential platform to study the strong interactions between single emitters and optical nanocavities. Future studies using emitters with a larger oscillator strength, such as quantum dots,[42] and exploring plasmonic nanoparticle cavities of different geometries could potentially enable light-matter interactions in strong coupling regime. Besides fundamental studies, this new platform will be useful to a wide variety of applications, including enhanced fluorescence imaging, single-photon sources, quantum computing, *etc*.

## METHODS

**Materials.** Ethylenediaminetetraacetic acid (EDTA, ACS grade), magnesium acetate tetrahydrate phenylbis (ACS grade), 2-amino-2-(hydroxymethyl)-1,3-propanediol (Tris base), sodium hydroxide (≥98%), boric acid (≥99.5%), gold(III) chloride trihydrate (≥99.9%), hexadecyltrimethylammonium bromide (CTAB, ≥99%), silver nitrate (ACS reagent, ≥99.0%), sodium borohydride (99%), L-Ascorbic acid, agarose powder for molecular biology and tris(2-carboxyethyl)phosphine hydrochloride (TCEP) were purchased from Sigma-Aldrich. Isopropyl alcohol, hydrochloric acid (36.5 to 38.0%), glacial acetic acid, Invitrogen™ 1 kb Plus DNA ladder and Thermo Scientific™ 6x orange DNA loading dye were purchased from Fisher Scientific. Uranyl formate (UF, 99-100%) was purchased from VWR. Carboxyethylsilanetriol disodium salt (CES, 25% in water) was purchased from Gelest. Fused silica was purchased from University Wafer. M13mp18 Single-stranded DNA was purchased from New England Biolabs. All the staple strands were custom-synthesized by Integrated DNA Technologies (IDT) and received in the form of 200 μM solutions in 1× TE buffer. All the other oligonucleotides were received in their lyophilized form from IDT. Their sequences are listed in Table S2.

**Buffers.** TAE buffer was prepared as a 50× stock solution. The 50× stock solution was prepared by dissolving 242 g Tris base in water, adding 57.1 mL glacial acetic acid, and 100 mL of 500 mM EDTA (pH 8.0) solution, and bringing the final volume up to 1 liter. The stock solution was diluted 50 folds for experimental use. The 1× TAE buffer (40 mM tris, 20 mM acetate, 1 mM EDTA) that had been previously diluted from 50× stock solution might also be supplemented with 12.5 mM magnesium acetate tetrahydrate, herein referred to as 1× TAE/$Mg^{2+}$. TBE buffer was prepared as a 10× stock solution. The stock solution was prepared by dissolving 108 g Tris base in water, adding 55 g of boric acid and 7.5 g EDTA, and bringing the final volume up to 1 liter. The stock solution was diluted to 1× for experimental use. Tris buffer was prepared by dissolving Tris base into DI water and adjusting the pH with 1 M HCl. The final concentration of Tris base was 10 mM with a pH of 8.3. The Tris-based stabilizing buffer was prepared in a similar manner. Tris base and magnesium acetate were first dissolved in DI water. The pH of the buffer was adjusted to 8.9. The final concentration of Tris and $Mg^{2+}$ was 10 mM and 15 mM, respectively.

**Preparation of DR.** The DR was designed using Cadnano to include two tubular grooves and a "spacer" according to the geometry of home-synthesized AuNRs (Figure S5). The overall dimension of DR was expected to be 58.3 nm (L) × 30.0 nm (W) × 15.0 nm (H), based on the known dimensions of DNA double helix: 3.4 nm long per 10.5 base pair (bp) and 2.5 nm wide per DNA helix (including the inter-helix gap).[18] A group of 3 docking strands (AuL1_12- AuR3_12, see Table S2) were extended from selected helper strands on each side of the DR. These docking strands were expected to bind two DNA-decorated gold nanorods in a linear configuration *via* DNA hybridization. Another docking strand (emitter in Table S2) was designed at the top center of the spacer to tether DNA labeled dye molecule. The DNA structure was obtained by folding a circular single-stranded M13mp18 DNA with a set of short helper strands at 1:10 molar ratio in 1× TAE/$Mg^{2+}$ buffer. The mixture solution was annealed in a thermocycler programmed for a cooling ramp from 90 °C to 25 °C over 12 hours. The DNA labeled fluorescent dye (Cy5) was also incorporated during the annealing process. The assembled DR was purified with a 100 kD Amicon filter under centrifugation at 8000 rpm for 10 min 3 times.

**Characterization of DR.** After purification, the correct intact structure of the DR was verified by gel electrophoresis (GE), AFM and TEM imaging while the concentration of the DR was estimated from the UV absorption at 260 nm. 1% agarose gel for GE tests was prepared by dissolving 1.00 g of agarose powder into 100 mL 1× TBE buffer containing 2 μL SYBR Safe DNA gel stain (Invitrogen). DO samples and DNA ladders were ran at 75 V for 35 min.

TEM images were taken with a Philips CM 12 microscope. Samples containing DNA origami were negatively stained with 1% UF for 30 s. AFM measurements were carried out with a Dimension FastScan system in the "ScanAsyst in fluid" mode using Scanasyst-Fluid+ tips (Bruker). For AFM imaging, 2 μL of the assembled sample was first diluted five times with 1× TAE/$Mg^{2+}$ buffer, then 2 μL diluted sample was deposited onto a freshly cleaved mica surface (Ted Pella), and 60 μL 1× TAE/$Mg^{2+}$ buffer was added on the top of the sample. After incubation for about 2 min to allow adsorption of the sample to the mica surface, the buffer solution was removed with a pipette. This also removed the most of the unbound helper strands to prevent their adverse effect on AFM imaging. Then 60 μL buffer was added to the sample again, and an additional 60 μL was deposited on the AFM tip before engaging.

**Synthesis of gold nanorod (AuNRs).** The AuNRs were synthesized following a previous work in our group with slightly modified recipes.[33] Briefly, 100 mM CTAB, 1 mM $AgNO_3$ and 79 mM ascorbic acid stock solutions were prepared first. $HAuCl_4$ was dissolved in 100 mM CTAB to yield a 1.5 mM $HAuCl_4$ stock solution. 60 μL ice-cold 0.010 M $NaBH_4$ was quickly added to 1 mL 0.25 mM $HAuCl_4$ 100 mM CTAB solution and vigorously stirred for 2 min. The solution should turn from dark yellow to brownish yellow immediately, which contained ~3.5 nm gold nanoparticles serving as seeds.[33] To synthesize AuNRs of various AR, desired amount of $AgNO_3$ solution, ascorbic acid solution and $HAuCl_4$ stock solution of varying volume ratios were mixed. To that mixture, a certain amount of the seed solution was added. The detailed recipe is listed in Table S3 and S4. The reaction mixture was then left undisturbed overnight to get the final products. The actual size of the as-synthesized AuNRs was characterized using a Philips CM 12 TEM. TEM images were analyzed by ImageJ.

**Surface DNA modification of AuNRs.** The AuNRs were modified with capping DNA sequences to pair with the docking DNA strands on the DO. The sequence of the capping DNA is listed in Table S2. To do the modification, dithiolated DNA was dissolved in DI water to yield a 1 mM solution. A 200 mM TCEP solution was prepared. The TCEP solution was mixed with DNA solution at equal volume and incubated at RT for 4 hours during which the dithiol bond was reduced to –SH. The as-prepared DNA-SH was then purified with Amicon ultra centrifugal filters (3k MW cutoff) at 15k rpm for 10 min to remove excess TCEP and small molecules, and the concentration of DNA-SH was determined by measuring its absorption at 260 nm using a NanoDrop spectrometer (Thermo Scientific).

The synthesized AuNRs were purified by centrifugation at 9000 rpm for 10 min four times to remove CTAB. The supernatant was discarded and the pellet was re-suspended in DI water. After the last spin, the pellet was dissolved in 1× TBE solution and 0.01% SDS was added. The concentration of the purified AuNR was determined by its maximum absorbance measured using UV-Vis spectrometer (Bruker) assuming that an absorbance of 1 equaled to 0.4 nM. 100 μL 1 mM thiol-modified DNA strand was mixed with 1 mM 1 nM AuNR and the mixture was left at RT overnight. After that, 5 M NaCl was gradually added to the above mixture within 48 hours to increase the salt concentration to 500 mM. The DNA modified AuNRs were purified by centrifugation at 8000 rpm, 10 min for four times to remove excessive DNA and salts. The supernatant was replaced by a 1× TAE buffer (without $Mg^{2+}$) after each spin. After the last centrifugation, the concentration of the product was measured by UV-Vis spectrometry and diluted to 1 nM with 1× TAE/$Mg^{2+}$ buffer.

**Preparation of AuNR-DR complexes.** AuNRs and DR were self-assembled under thermal annealing. Fluorescence thermal curves were measured in optical tube strips using an MX3005P real-time thermocycler (Agilent Technology) equipped with a fluorescence 96-well plate reader. The DR was mixed with 1× SYBR Green I (Invitrogen) in 1× TAE/$Mg^{2+}$ buffer. The fluorescence intensity of the emission was monitored at 522 nm with excitation at 495 nm at 1 min intervals. The samples were first heated to 85 °C for 10 min, and the temperature was reduced from 85 to 25 °C at a rate of -0.5 °C/min. After cooling down to 25 °C, the samples were held for 10 min and then heated to 85 °C with a temperature gradient of +0.5 °C/min. The thermal curve of AuNR capping strands binding with their complementary strands was measured using the same setup. The DNA-capped AuNRs and the DR were mixed at a 3:1 molar ratio in 1× TAE/$Mg^{2+}$ buffer. The mixture was then annealed from 45 to 30 °C at a rate of –0.02 °C/min. The annealing process was repeated 4 times to increase the final yield. For comparison, DR bound AuNR monomers were also prepared by replacing the docking strands on one side (AuR1_12, AuR2_12, AuR3_12 in Table S2) with the blocking strands (BR1_12, BR2_12, BR3_12 in Table S2). The as-modified DR was only able to capture one AuNR inside one of its two grooves.

**Modification of silica substrate.** To enhance the affinity between the AuNR-DR complex and fused silica substrate, the surface of the silica was chemically modified with a layer of carboxyl groups. Fused silica chips were first cleaned by sonication (Branson 5800) in IPA and DI water for 5 min, respectively. This step was repeated twice to remove most physically absorbed organic and inorganic contaminants. Next, the chips were dried in air and cleaned with 270 W oxygen plasma (Harrick Plasma expanded plasma cleaner) for 180 s. The chips were then immersed in the Tris buffer (pH 8.3) containing 0.01% CES for 10 min, followed by sonication in DI water.

**Immobilization of AuNR-DO complexes.** A 50 mm Petri dish was prepared with a moistened piece of laboratory tissue paper to limit evaporation. Solution with 100 pM AuNR-DR complexes was prepared in Tris buffer, and a 20 μL drop was deposited in the middle of a fused silica chip. The chip was placed in the Petri dish, and the sample solution was allowed to incubate on the chip for 1 h with the Petri dish covered. After the incubation, excess sample in solution was washed away by 60 μL of fresh Tris buffer pipetted onto and taken off the chip, repeated 8 times. Next, the chip was again washed 8 times with the stabilizing buffer in order to allow the DO to bind strongly and to minimize artifacts during subsequent drying. The chip was finally dipped in 50% ethanol (v/v% in water) for 10 s, 75% ethanol for 10 s, and 90% ethanol for 120 s. Then the chip was air-dried and ready for further tests.

**Microscopic and spectral characterizations.** The microscopic bright field imaging, dark field imaging, spectra collection, and lifetime analysis were performed using a customized optical system (Figure S10). An upright Olympus BX53 microscope was equipped with a Xeon lamp (PowerArc, Horiba). The power density was measured to be 4.0 W/cm$^2$ after the objective. The sample-coated silica chips were illuminated, and bright filed images were recorded with a charge-coupled device (CCD) camera. Dark field images were taken using an EMCCD camera (iXon Ultra, Andor) for enhanced sensitivity through 100x dark field lens (numerical aperture NA=0.9). The detector was thermoelectrically cooled to -100 °C while the electron multiplication gain was set at 64. Spectral measurements were taken using a microscope-coupled UV-Vis-NIR spectrometer (Horiba iHR 320) equipped with a CCD detector. Fluorescence lifetime was measured by a PPD 900 photo counting detector (Horiba) using SuperK EVO supercontinuum fiber laser (NKT Photonics) as the light source. The power density was measured to be 12.1 W/cm$^2$ after the objective. The integration was manually stopped when the measured peak height exceeded 1000 counts, which commonly took 10-15 min for the surface immobilized emitters, and 3-4 hours for the prompt signals. The region of interest on a sample was first found under bright field and dark field imaging mode. The dark field spectral data was then taken from the same region, followed by fluorescence measurements and lifetime measurements. Note that Cy5 might undergo photobleaching after continuous strong illuminations so the above optical measurements of a particular region was only conducted once. Lifetime value was extracted from the raw data using the DAS6 software from Horiba.

**FDTD simulations.** FDTD simulations were conducted with the commercial Ansys Lumerical software. The AuNR dimer is set with a width of 12 nm while the length varied in different simulations. The diameter of AuNS was set in such a way that its volume equaled to that of the AuNR in a parallel comparison. The gap between the AuNS or AuNR dimers was set as 4 nm. The complex refractive index (ñ) data of gold was defined following the experiment measured data of crystalline gold in literature.[53] The fused silica substrate material was modeled as Palik.[53] The ñ of the background was set as 1. The docking DNA origami was modeled as a dielectric material with a constant ñ of 1.53. The docking DNA origami was 58 nm long, 30 nm wide and 6 nm tall. The Cy5 or Cy7 dye was modeled as a cube whose real part of ñ (n) was 1.54 and the imaginary part (κ) was defined based on the absorption spectra of Cy5 and Cy7, respectively.[51,54] Note the dye was a single molecule, whose extinction coefficient should differ from that of the bulk solution.

Here the maximum value of κ was set as 0.4. The effect of κ on the final scattering spectra is plotted in Figure S20. The FDTD boundary condition was set as perfect matched layers (PML) in all directions (8 layers). The mesh size was set as 0.5 nm in x, y and z directions in the area where AuNS or AuNR and the dye were located. Total field scattered field (TFSF) source was employed for scattering simulations. Five monitor planes were placed outside the TFSF source zone and covered both x, y directions and positive z direction of the emitters in order to capture the scattering spectra.

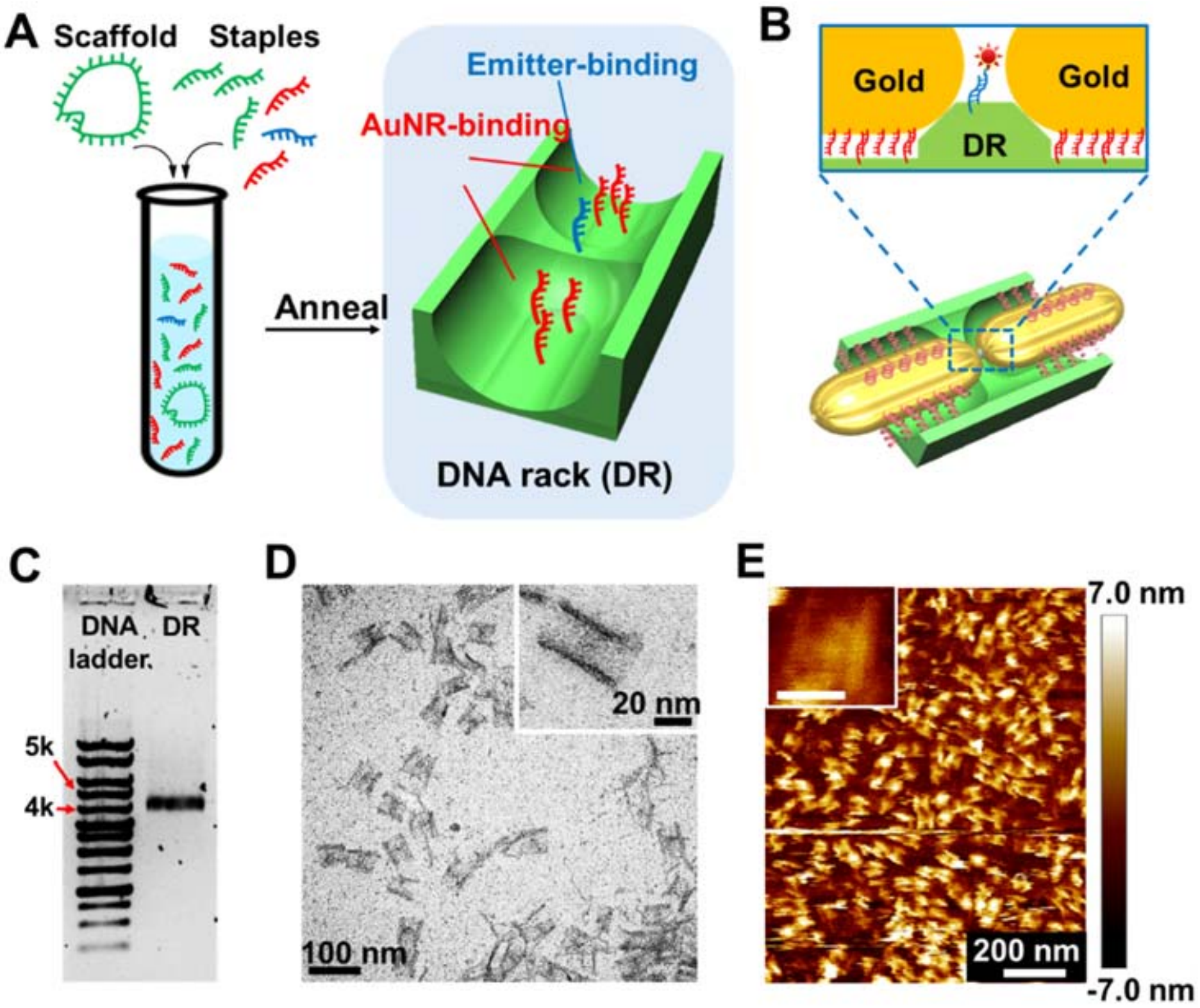

**Figure 1**. Design of DR and its application in constructing plasmonic nanocavities. (A) Simple scheme of the synthesis of DR. (B) 3D scheme of the design of anisotropic plasmonic optical cavity with a single emitter. Inset: side view of the nanocavity. (C) Agarose gel electrophoresis image of the purified DR samples. (D) TEM image of DR. Inset: a magnified single DR. (E) AFM image of DR. Inset: a magnified image of a single DR. Scale bar: 30 nm.

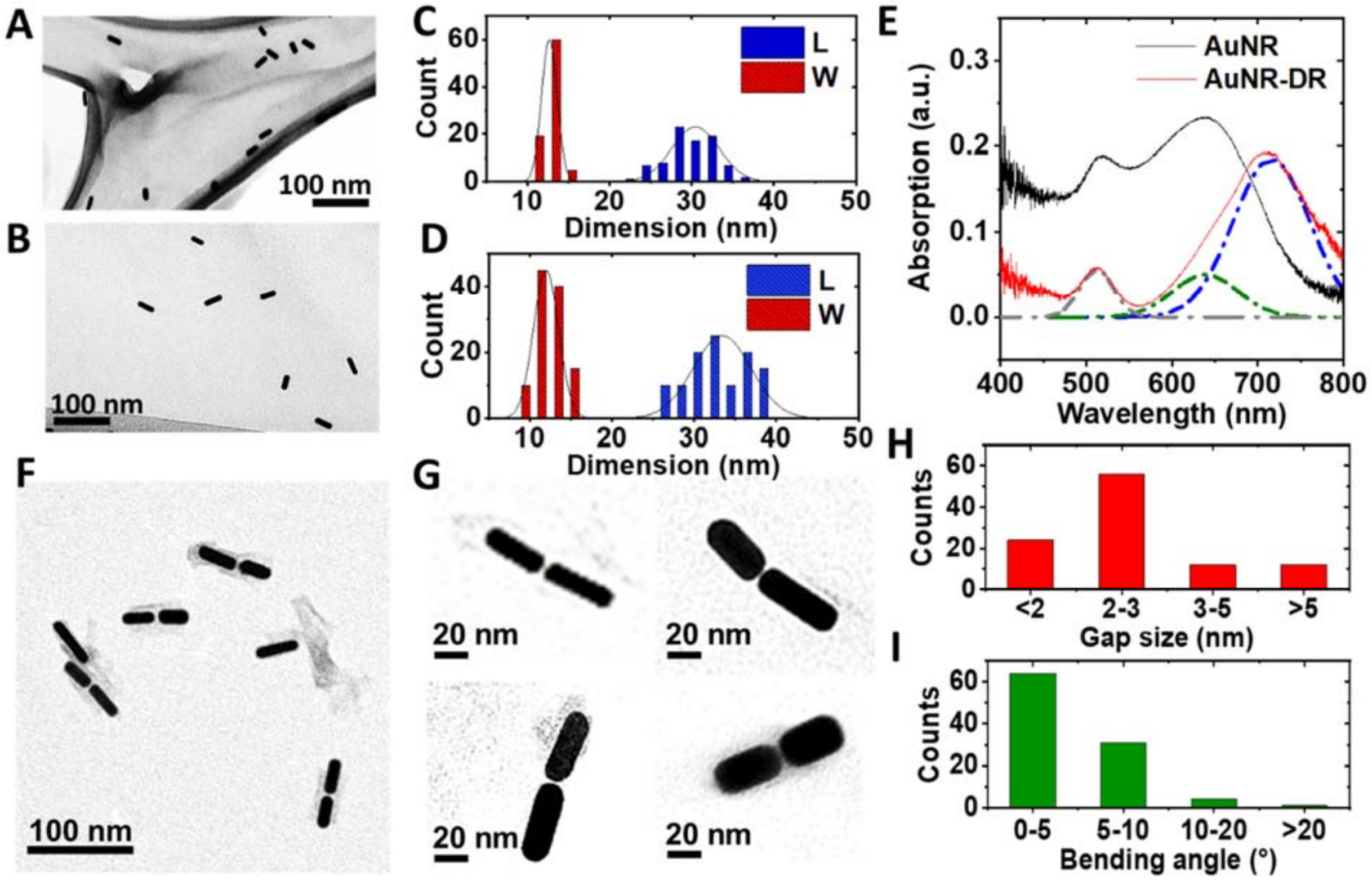

**Figure 2**. Structural characterization of the fabricated plasmonic AuNR nanocavities via DO guided self-assembly. (A) TEM image of home-synthesized AuNRs (AR=2.5). (B) TEM image of home-synthesized AuNRs (AR=2.8). (C) Size distributions of the AuNRs in (A). (C) Size distributions of the AuNRs in (B). (E) UV-Vis spectra of AuNR monomers and AuNR dimer-DR complexes. The green and blue dashed lines correspond to the peak fitting result for the longitudinal plasmonic mode. (F) TEM image of fabricated AuNR-DR complexes. (G) Magnified TEM images of plasmonic nanocavity. (H) Distribution of the measured gap size in the nanocavities. (I) Distribution of the measured bending angle of the AuNR dimers.

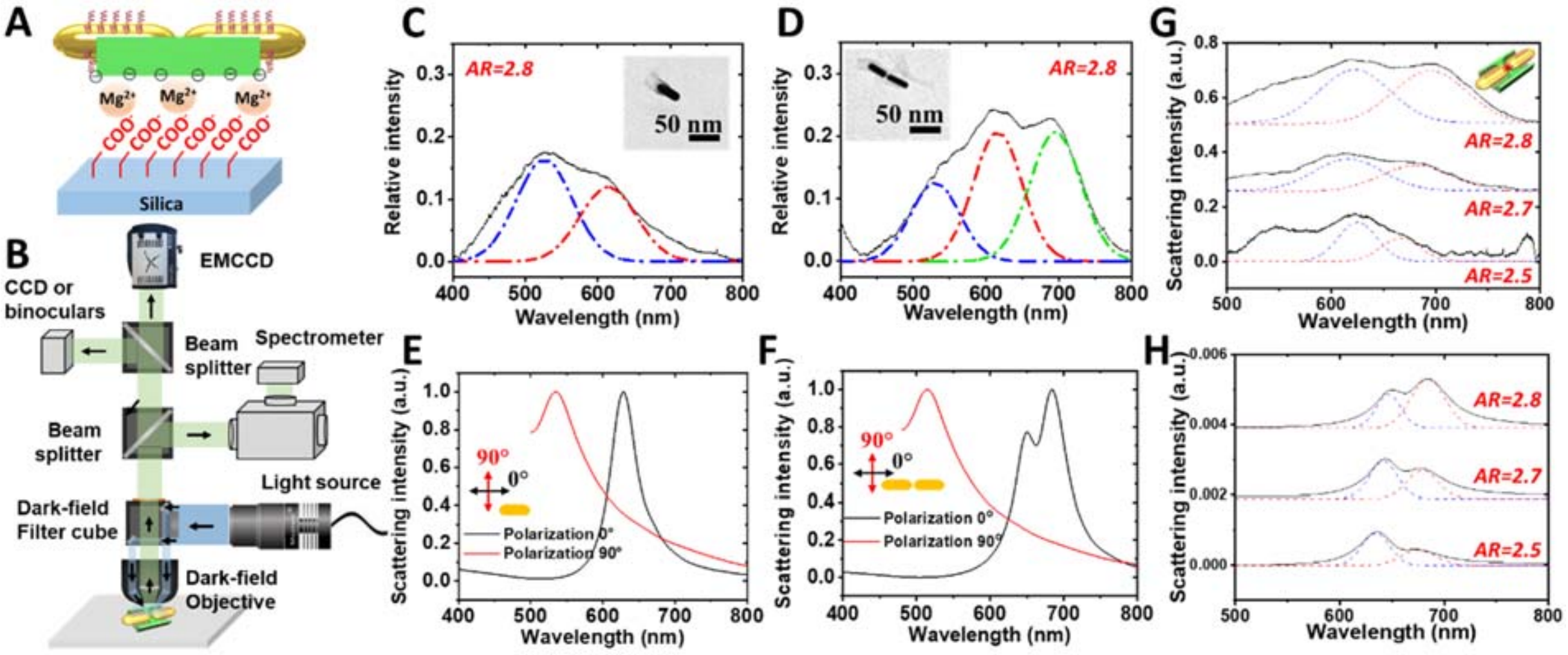

**Figure 3**. Dark field scattering characterizations of surface immobilized AuNR-DR complexes with assembled single emitters. (A) Scheme of the substrate immobilization mechanism. (B) Scheme of the experimental setup. Scattered light was collected by a spectrometer for spectral analysis. (C-D) DFS spectrum of the surface immobilized DR bound AuNR monomers (panel C) and dimers (panel D). The dash-dotted lines are the peak fitting results. Inset: representative TEM images of a DR bound AuNR monomer and dimer. The AuNRs are 12 nm in width and 30 nm long. (E-F) FDTD-calculated polarization-dependent DFS spectra of DR bound AuNR monomers (panel E) and dimers (panel F). The dimensions of AuNRs are the same as used in experiments (panels C and D). (G) The experimentally measured DFS spectra of plasmonic emitters made of AuNR dimers of various ARs. (H) Simulated DFS spectra of plasmonic quantum emitters made of AuNR dimers of various ARs.

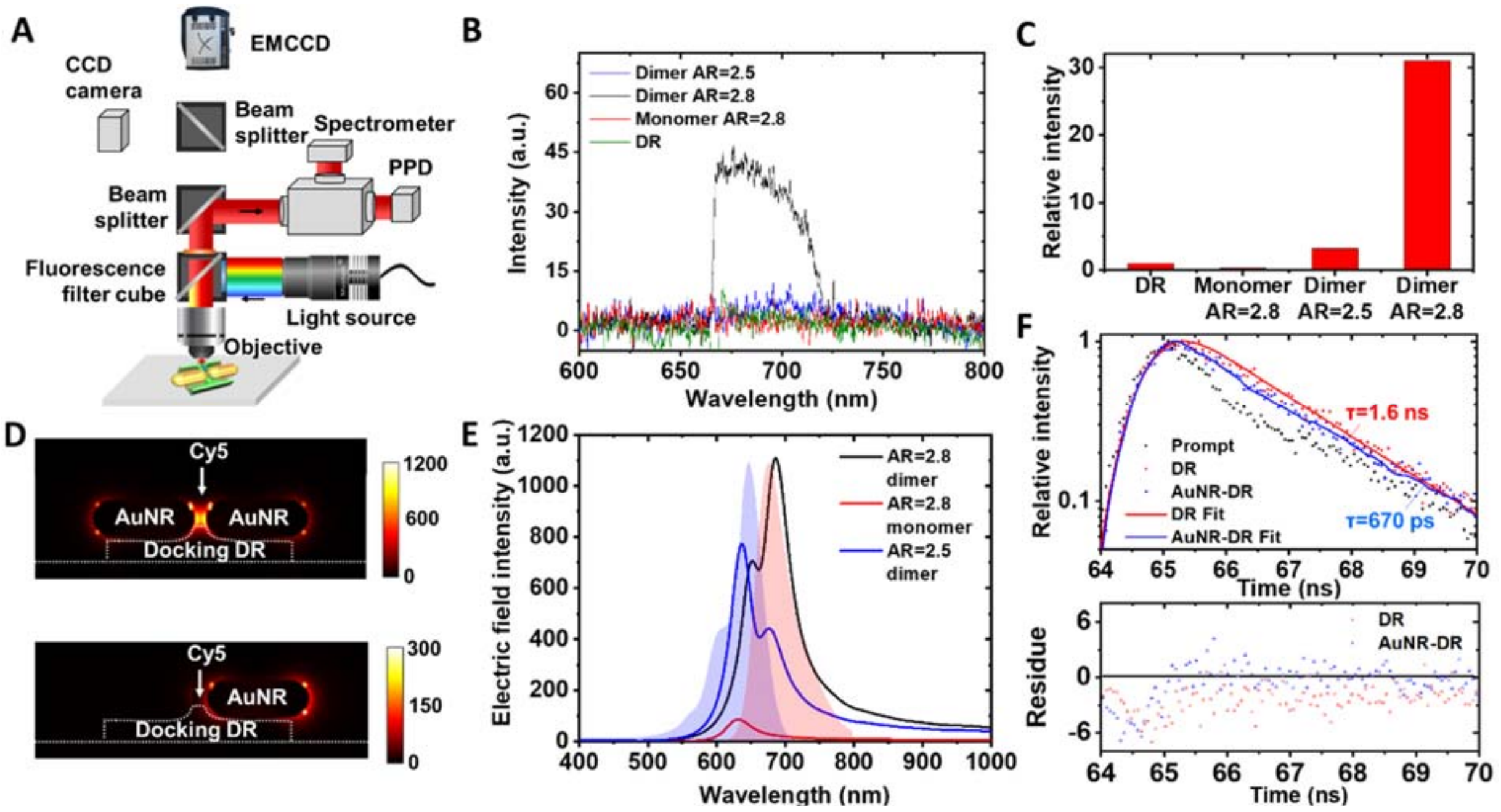

**Figure 4**. Fluorescence characterizations of the single emitter assembled in plasmonic nanocavities. (A) Scheme of the experimental setup. Fluorescence signals were collected by a spectrometer for spectral analysis or a PPD for lifetime measurements. (B) The measured fluorescence emission spectra of various samples. (C) Normalized fluorescence intensity of the samples in (B). (D) Simulated EF distribution in the vicinity of AR=2.8 AuNR dimers (top) and AuNR monomers (bottom). The hotspots are clearly visible at the center spots of the AuNR dimer or close to the tips of the AuNR monomer. (E) Simulated electric field intensity as a function of wavelength at the nanocavity for both AuNR dimer and monomer. The Cy5 absorption and emission spectra were shown as shaded areas for comparison. (F) Fluorescence lifetime measurements and the corresponding residue of Cy5 with and without the plasmonic nanocavity.

ACKNOWLEDGMENT

Y.L. thanks the support from an Army Research Office MURI award no. W911NF-12-1-0420. C.W. thanks the ASU startup funds and National Science Foundation under award no. 1711412, 1838443 and 1847324 for partially supporting this research. Y.Y. thanks the ASU startup funds and National Science Foundation under award no. 1809997 for partially supporting this research.

**Electronic Supplementary Material**

The following files are available in the online version of this article.

Additional details on the experimental conditions; mathematical analysis on the fluorescence lifetime and Purcell factor; additional sample characterizations (PDF)

**Notes**

The authors declare no competing financial interest.

REFERENCES

[1] Tame, M. S.; McEnery, K. R.; Özdemir, Ş. K.; Lee, J.; Maier, S. A.; Kim, M. S. Quantum plasmonics. *Nat. Phys.* **2013**, *9*, 329-340.

[2] Thompson, J. D.; Tiecke, T. G.; de Leon, N. P.; Feist, J.; Akimov, A. V.; Gullans, M.; Zibrov, A. S.; Vuletić, V.; Lukin, M. D. Coupling a single trapped atom to a nanoscale optical cavity. *Science* **2013**, *340*, 1202-1205.

[3] Faraon, A.; Fushman, I.; Englund, D.; Stoltz, N.; Petroff, P.; Vučković, J. Coherent generation of non-classical light on a chip via photon-induced tunnelling and blockade. *Nat. Phys.* **2008**, *4*, 859-863.

[4] Andreani, L. C.; Panzarini, G.; Gérard, J. M. Strong-coupling regime for quantum boxes in pillar microcavities: theory. *Phys. Rev. B* **1999**, *60*, 13276.

[5] Minder, M.; Pittaluga, M.; Roberts, G. L.; Lucamarini, M.; Dynes, J. F.; Yuan, Z. L.; Shields, A. J. Experimental quantum key distribution beyond the repeaterless secret key capacity. *Nat. Photon.* **2019**, *13*, 334-338.

[6] Hennessy, K.; Badolato, A.; Winger, M.; Gerace, D.; Atatüre, M.; Gulde, S.; Fält, S.; Hu, E. L.; Imamoğlu, A. Quantum nature of a strongly coupled single quantum dot-cavity system. *Nature* **2007**, *445*, 896-899.

[7] Schlather, A. E.; Large, N.; Urban, A. S.; Nordlander, P.; Halas, N. J. Near-field mediated plexcitonic coupling and giant Rabi splitting in individual metallic dimers. *Nano Lett.* **2013**, *13*, 3281-3286.

[8] Reithmaier, J. P.; Sęk, G.; Löffler, A.; Hofmann, C.; Kuhn, S.; Reitzenstein, S.; Keldysh, L. V.; Kulakovskii, V. D.; Reinecke, T. L.; Forchel, A. Strong coupling in a single quantum dot-semiconductor microcavity system. *Nature* **2004**, *432*, 197-200.

[9] Srinivasan, K.; Painter, O. Linear and nonlinear optical spectroscopy of a strongly coupled microdisk–quantum dot system. *Nature* **2007**, *450*, 862-865.

[10] Thon, S. M.; Rakher, M. T.; Kim, H.; Gudat, J.; Irvine, W. T.; Petroff, P. M.; Bouwmeester, D. Strong coupling through optical positioning of a quantum dot in a photonic crystal cavity. *Appl. Phys. Lett.* **2009**, *94*, 111115.

[11] Gopinath, A.; Miyazono, E.; Faraon, A.; Rothemund, P. W. Engineering and mapping nanocavity emission via precision placement of DNA origami. *Nature* **2016**, *535*, 401-405.

[12] Akimov, A. V.; Mukherjee, A.; Yu, C. L.; Chang, D. E.; Zibrov, A. S.; Hemmer, P. R.; Park, H.; Lukin, M. D. Generation of single optical plasmons in metallic nanowires coupled to quantum dots. *Nature* **2007**, *450*, 402-406.

[13] Chikkaraddy, R.; de Nijs, B.; Benz, F.; Barrow, S. J.; Scherman, O. A.; Rosta, E.; Demetriadou, A.; Fox, P.; Hess, O.; Baumberg, J. J. Single-molecule strong coupling at room temperature in plasmonic nanocavities. *Nature* **2016**, *535*, 127-130.

[14] Hoang, T. B.; Akselrod, G. M.; Argyropoulos, C.; Huang, J.; Smith, D. R.; Mikkelsen, M. H. Ultrafast spontaneous emission source using plasmonic nanoantennas. *Nat. Commun.* **2015**, *6*, 7788.

[15] Van der Sar, T.; Hagemeier, J.; Pfaff, W.; Heeres, E. C.; Thon, S. M.; Kim, H.; Petroff, P. M.; Oosterkamp, T. H.; Bouwmeester, D.; Hanson, R. Deterministic nanoassembly of a coupled quantum emitter-photonic crystal cavity system. *Appl. Phys. Lett.*, **2011**, *98*, 193103.

[16] Riedrich-Möller, J.; Arend, C.; Pauly, C.; Mücklich, F.; Fischer, M.; Gsell, S.; Schreck, M.; Becher, C. Deterministic coupling of a single silicon-vacancy color center to a photonic crystal cavity in diamond. *Nano Lett.* **2014**, *14*, 5281-5287.

[17] Rothemund, P. W. K. Folding DNA to create nanoscale shapes and patterns. *Nature* **2006**, *440*, 297-302.

[18] Han, D.; Pal, S.; Nangreave, J.; Deng, Z.; Liu, Y.; Yan, H. DNA origami with complex curvatures in three-dimensional space. *Science* **2011**, *332*, 342-346.

[19] Samanta, A.; Zhou, Y.; Zou, S.; Yan, H.; Y. Liu Fluorescence quenching of quantum dots by gold nanoparticles: a potential long range spectroscopic ruler. *Nano Lett.* **2014**, *14*, 5052–5057.

[20] Pal, S.; Dutta, P.; Wang, H.; Deng, Z.; Zou, S.; Yan, H.; Liu, Y. Quantum efficiency modification of organic fluorophores using gold nanoparticles on DNA origami scaffolds. *J. Phys. Chem. C* **2013**, *117*, 12735-12744.

[21] Xin, L.; Lu, M.; Both, S.; Pfeiffer, M.; Urban, M. J.; Zhou, C.; Yan, H.; Weiss, T.; Liu, N.; Lindfors, K. Watching a single fluorophore molecule walk into a plasmonic hotspot. *ACS Photon.* **2019**, *6*, 985-993.

[22] Vietz, C.; Kaminska, I.; Sanz Paz, M.; Tinnefeld, P.; Acuna, G. P. Broadband fluorescence enhancement with self-assembled silver nanoparticle optical antennas. *ACS Nano* **2017**, *11*, 4969-4975.

[23] Chikkaraddy, R.; Turek, V. A.; Kongsuwan, N.; Benz, F.; Carnegie, C.; van de Goor, T.; de Nijs, B.; Demetriadou, A.; Hess, O.; Keyser, U. F. Baumberg, J. J. Mapping nanoscale hotspots with single-molecule emitters assembled into plasmonic nanocavities using DNA origami. *Nano Lett.* **2017**, *18*, 405-411.

[24] Roller, E. M.; Argyropoulos, C.; Högele, A.; Liedl, T.; Pilo-Pais, M. Plasmon-exciton coupling using DNA templates. *Nano Lett.* **2016**, *16*, 5962-5966.

[25] Lin, K. Q.; Yi, J.; Hu, S.; Liu, B. J.; Liu, J. Y.; Wang, X.; Ren, B. Size effect on SERS of gold nanorods demonstrated via single nanoparticle spectroscopy. *J. Phys. Chem. C* **2016**, *120*, 20806-20813.

[26] Huang, C. P.; Yin, X. G.; Kong, L. B.; Zhu, Y. Y. Interactions of nanorod particles in the strong coupling regime. *J. Phys. Chem. C* **2010**, *114*, 21123-21131.

[27] Jain, P. K.; Eustis, S.; El-Sayed, M. A. Plasmon coupling in nanorod assemblies: optical absorption, discrete dipole approximation simulation, and exciton-coupling model. *J. Phys. Chem. B* **2006**, *110*, 18243-18253.

[28] Funston, A. M.; Novo, C.; Davis, T. J.; Mulvaney, P. Plasmon coupling of gold nanorods at short distances and in different geometries. *Nano Lett.* **2009**, *9*, 1651-1658.

[29] Shao, L.; Woo, K. C.; Chen, H.; Jin, Z.; Wang, J.; Lin, H. Q. Angle-and energy-resolved plasmon coupling in gold nanorod dimers. *ACS Nano* **2010**, *4*, 3053-3062.

[30] Link, S.; Mohamed, M. B.; El-Sayed, M. A. Simulation of the optical absorption spectra of gold nanorods as a function of their aspect ratio and the effect of the medium dielectric constant. *J. Phys. Chem. B* **1999**, *103*, 3073-3077.

[31] Süel, G. Use of fluorescence microscopy to analyze genetic circuit dynamics. In *Methods in Enzymology*. Academic Press: New York, 2011; pp 275-293.

[32] Liu, X.; Zhang, F.; Jing, X.; Pan, M.; Liu, P.; Li, W.; Zhu, B.; Li, J.; Chen, H.; Wang, L.; Lin, J.; Liu, Y.; Zhao, D.; Yan, H.; Fan, C. Complex silica composite nanomaterials templated with DNA origami. *Nature* **2018**, *559*, 593-598.

[33] Gole, A.; Murphy, C. J. Seed-mediated synthesis of gold nanorods: role of the size and nature of the seed. *Chem. Mater.* **2004**, *16*, 3633-3640.

[34] Mirkin, C. A.; Letsinger, R. L.; Mucic, R. C.; Storhoff, J. J. A DNA-based method for rationally assembling nanoparticles into macroscopic materials. *Nature* **1996**, *382*, 607-609.

[35] Cocco, S.; Marko, J. F.; Monasson, R. Theoretical models for single-molecule DNA and RNA experiments: from elasticity to unzipping. *C R Phys.* **2002**, *3*, 569-584.

[36] Roth, E.; Glick Azaria, A.; Girshevitz, O.; Bitler, A.; Garini, Y. Measuring the conformation and persistence length of single-stranded DNA using a DNA origami structure. *Nano Lett.* **2018**, *18*, 6703-6709.

[37] Chi, Q.; Wang, G.; Jiang, J. The persistence length and length per base of single-stranded DNA obtained from fluorescence correlation spectroscopy measurements using mean field theory. *Physica A Stat. Mech. Appl.* **2013**, *392*, 1072-1079.

[38] Hagerman, P. J. Flexibility of DNA. *Annu. Rev. Biophys. Biophys. Chem.* **1988**, *17*, 265-286.

[39] Thacker, V. V.; Herrmann, L. O.; Sigle, D. O.; Zhang, T.; Liedl, T.; Baumberg, J. J.; Keyser, U. F. DNA origami based assembly of gold nanoparticle dimers for surface-enhanced Raman scattering. *Nat. Commun.* **2014**, *5*, 3448.

[40] Simoncelli, S.; Roller, E. M.; Urban, P.; Schreiber, R.; Turberfield, A. J.; Liedl, T.; Lohmüller, T. Quantitative single-molecule surface-enhanced Raman scattering by optothermal tuning of DNA origami-assembled plasmonic nanoantennas. *ACS Nano* **2016**, *10*, 9809-9815.

[41] Chang, W. S.; Ha, J. W.; Slaughter, L. S.; Link, S. Plasmonic nanorod absorbers as orientation sensors. *Proc. Natl. Acad. Sci.* **2010**, *107*, 2781-2786.

[42] Moreels, I.; Lambert, K.; De Muynck, D.; Vanhaecke, F.; Poelman, D.; Martins, J. C.; Allan, G.; Hens, Z. Composition and size-dependent extinction coefficient of colloidal PbSe quantum dots. *Chem. Mater.* **2007**, *19*, 6101-6106.

[43] Cai, Y. Y.; Liu, J. G.; Tauzin, L. J.; Huang, D.; Sung, E.; Zhang, H.; Joplin, A.; Chang, W. S.; Nordlander, P.; Link, S. Photoluminescence of gold nanorods: Purcell effect enhanced emission from hot carriers. *ACS Nano* **2018**, *12*, 976-985.

[44] Fort, E.; Grésillon, S. Surface enhanced fluorescence. *J. Phys. D: Appl. Phys.* **2007**, *41*, 013001.

[45] Giannini, V.; Fernández-Domínguez, A. I.; Heck, S. C.; Maier, S. A. Plasmonic nanoantennas: fundamentals and their use in controlling the radiative properties of nanoemitters. *Chem. Rev.* **2011**, *111*, 3888-3912.

[46] Purcell, E. M. Spontaneous emission probabilities at radio frequencies. In *Confined Electrons and Photons*. Springer: Boston, 1995; pp 839.

[47] Kinkhabwala, A.; Yu, Z.; Fan, S.; Avlasevich, Y.; Müllen, K.; Moerner, W. E. Large single-molecule fluorescence enhancements produced by a bowtie nanoantenna. *Nat. Photon.* **2009**, *3*, 654-657.

[48] Rose, A.; Hoang, T. B.; McGuire, F.; Mock, J. J.; Ciracì, C.; Smith, D. R.; Mikkelsen, M. H. Control of radiative processes using tunable plasmonic nanopatch antennas. *Nano Lett.* **2014**, *14*, 4797-4802.

[49] Geddes, C. D.; Lakowicz, J. R. Metal-enhanced fluorescence. *J. Fluoresc.* **2002**, *12*, 121-129.

[50] Malicka, J.; Gryczynski, I.; Fang, J.; Kusba, J.; Lakowicz, J. R. Photostability of Cy3 and Cy5-labeled DNA in the presence of metallic silver particles. *J. Fluoresc.* **2002**, *12*, 439-447.

[51] Yoshie, T.; Scherer, A.; Hendrickson, J.; Khitrova, G.; Gibbs, H. M.; Rupper, G.; Ell, C.; Shchekin, O. B.; Deppe, D. G. Vacuum Rabi splitting with a single quantum dot in a photonic crystal nanocavity. *Nature* **2004**, *432*, 200-203.

[52] Moreira, B. G.; You, Y.; Owczarzy, R. Cy3 and Cy5 dyes attached to oligonucleotide terminus stabilize DNA duplexes: predictive thermodynamic model. *Biophys. Chem.* **2015**, *198*, 36-44.

[53] Olmon, R. L.; Slovick, B.; Johnson, T. W.; Shelton, D.; Oh, S.; Boreman, G. D., Raschke, M. B. Optical dielectric function of gold. *Phys. Rev. B* **2012**, *86*, 235147.

[54] Carmichael, H. J.; Brecha, R. J.; Raizen, M. G.; Kimble, H. J.; Rice, P. R. Subnatural linewidth averaging for coupled atomic and cavity-mode oscillators. *Phys. Rev. A* **1989**, *40*, 5516-5519.

# Supplementary information for

## Deterministic Assembly of Single Emitters in Sub-5 Nanometer Optical Cavity Formed by Gold Nanorod Dimers on Three-Dimensional DNA Origami

Zhi Zhao, Xiahui Chen, Jiawei Zuo, Ali Basiri, Shinhyuk Choi, Yu Yao, Yan Liu,* Chao Wang*

Correspondence to: Yan_Liu@asu.edu; wangch@asu.edu

**This file includes:**

Supplementary Text

Figs. S1 to S20

Table S1 to S4

**Supplementary text**

1. The gap sizes in plasmonic nanocavities assembled by DO templated methods previously reported.

Previously plasmonic nanocavities assembled using DO templates shared a similar design, i.e., two plasmonic structures were placed on the opposite sides of a DO via hybridization of the surface-labeling strands on the nanoparticles with the docking strands extended on the DO (Figure S1).[1,2] In such cases, the minimal cavity size was limited by the intrinsic thickness of the DO and the length of the DNA duplexes on both sides of the DO. For example, each layer of DO is estimated to be ~2.5 nm in thickness.[3] Meanwhile, the duplex regions for binding with the docking strands were usually longer than 10 base pairs (bps) per strand, which allowed their melting temperature to be significantly greater than room temperature.[4] In turn, this added at least 3.4 nm per side to the gap size of the nanocavities. As a result, the theoretical gap size of the nanocavities between the nanoparticles was minimally 2.5+3.4+3.4 = 9.3 nm. Even if some structural deformation takes place, the real gap sizes in the previously published literatures were unlikely to be smaller than 5 nm.[1,2]

2. Design principle of DR

Here in this report DR was designed in such a way that it was possible to guide the assembly of AuNR monomer or linear dimer and a single emitter via steric effect and specific DNA hybridization (Figure S2). We started from a planar DO with six docking strands, which was employed in assembling nanoparticles.[1,5] In order to properly align the AuNR-dimer into a linear configuration, sidewalls were added to limit the in-plane rotation of the AuNRs. Next, a spacer was added to guide the assembly of AuNRs and emitters (indicated by the red star). The spacer ensured that the two grooves would be occupied by two AuNRs at a well-controlled distance, instead of one AuNR in the center, and the emitter could be inserted in the top-center position of the spacer through DNA hybridization. Finally, the distribution of the docking strands was adjusted to produce ultra-small gaps. It was the width of the spacer that defined the smallest gap size of the AuNR dimer. In order to eliminate the formation of any larger gaps, the docking strands were designed in such a way that two thirds of the strands were placed close to the spacer. Consequently, both AuNRs tended to stay close to the center, which allowed them to bind the maximal number of docking strands and reach the most energetically favorable configuration.

3. Tuning the geometry of AuNRs by adjusting synthetic conditions.

AuNRs were synthesized via seed mediated AuNR growth. The seed solution was prepared by adding 60 μL ice-cold 0.010 M $NaBH_4$ to 1 mL 0.25 mM $HAuCl_4$-100 mM CTAB solution and vigorously stirring for 2 min. The geometry of synthesized AuNRs could be readily tuned via adjusting the concentration of $HAuCl_4$ and $AgNO_3$ (Figure S6). For example, the diameter of AuNRs increased with elevated Au(III) concentrations while keeping the other experimental conditions identical (Table S3). Note that Au(III) was dissolved in 100 mM CTAB and CTAB was in large excess. Thus the minimal variation in CTAB concentrations across different reaction mixtures could be neglected. Meanwhile, the AR of AuNRs correlated with the volume of $AgNO_3$ solution ($V_{Ag}$) added to the synthetic system (Table S4). When the $HAuCl_4$ concentration was fixed at 0.4 mM, the measured AR of AuNRs increased from 2.0 to 3.0 as the $V_{Ag}$ changed from 40 μL to 100 μL. This could be verified by the shift in SPR absorption of different samples. The successful preparation of AuNRs with various AR was confirmed by TEM imaging (Figure S7A-S7D). The length (L) and width (W) distributions of various AuNRs extracted from TEM images were analyzed statistically. Two representative results are shown in Figures S7E and S7F. It is evident that the width of AuNRs were well controlled at ~12 nm with small deviations (<0.5 nm). The length distribution varied from sample to sample, leading to AuNRs with a series of different AR. The narrow distribution of length made it possible to distinguish ~0.1 difference in AR directly from TEM images.

4. Fluorescent melting curves

Fluorescent melting curves (Figure S9) were measured using 8-well optical tube strips in an MX3005P real-time thermocycler (Strategene) following a previously reported protocol.[6] The change of fluorescent emission from a DNA intercalating dye (SYBR Green I) served as an indicator of DNA hybridization/melting. The initial concentration of DR was 1 nM and that of the DNA capping strand (with its complementary strand) was 1 μM. The temperature was first elevated from 25 °C to 85 °C at +0.5 °C/min, then cooled down to RT at -0.5 °C/min. The fluorescent signal from SYBR Green I was recorded at different temperatures.

5. Setup of the microscope-coupled UV-Vis spectrometer

The setup of the microscope-coupled UV-Vis spectrometer is home-built and schemed in Figure S10A. The system is mounted on an optical table and enclosed in a laser curtain. A xenon arc lamp (PTI power arc, Horiba) was used as the light source in the bright field mode, the dark field mode and the fluorescent spectra collection. SuperK EVO supercontinuum fiber laser (NKT Photonics) was utilized as the light source in lifetime measurements, and was coupled to the microscope through a series of mirrors and a periscope. A Horiba iHR 320 UV-Vis-NIR spectrometer was coupled to an Olympus BX53 microscope to collect spectral data as well as for lifetime measurement. The spectra data were captured by a Syncerity BI-NIR CCD camera (Horiba) and analyzed by Labspec software (Horiba). The fluorescent images were captured and analyzed by ImageJ. The lifetime measurement was carried out by coupling the signals deflected from the gratings to a photodetector (PPD 900, Horiba). The SuperK "Analog output" port is connected to the PPD TTL-Out-NIM port, and the data were captured and analyzed by software DAS6 (Horiba).

A few sets of filter cubes (bright-field, dark-field, and fluorescent) and objective lens (10X, 50X, and 100X dark-field) were selected to fulfill the purpose of various experimental modes (Figure S10B). The objectives were mounted on a 6-position turret compatible with dark-field signal collection. For example, a filter cube containing a 50/50 dichroic mirror (part no. 21000, Chroma Technology) was used for bright field mode for focusing on the sample, typically working with 10X and 50X objectives. For dark-field scattering signal collection, a specialized dark mode cube (part no. U-M703, Olympus) was employed in dark field scattering measurements, in combination with a 100X dark-field objective (Figure 10C). Using the light stopper on the cube, a ring-shaped light field was generated and illuminated on the sample. The scattered light emitted in all the directions. Only the portion that passed through the central lens in the 100X dark-field objective would be collected. In this way it guarantees that no incident light would interference with the collected scattering beam. In fluorescent mode, we chose filter cubes (part no. 39007, Chroma Technology) consisting of two filters and a longpass dichroic mirror. The excitation filter is a bandpass filter (595 nm-645 nm), the dichroic mirror is a longpass filter (>655 nm), and the emission filter is another bandpass filter (670-720 nm). When white light hit the excitation filter, only the selected band passed (Figure 10D). The light was then focused on the sample and excited the fluorescent emission. The emitted light had a different wavelength, which could pass the dichroic mirror and emission filter. No incident light could pass these two filters.

6. Understanding of light-matter interaction

We consider a single quantum emitter in a 3D optical cavity (Figure S14A),[7-10] where the emitter is modeled by a two-level system with a transition frequency $\omega_0$ and an isotropic dipole $d$. The coupling between the emitter and the cavity, with the coupling strength $g$ is defined as:

$$g = \left(\frac{\omega_0}{2\epsilon_r\epsilon_0 V_{eff}\hbar}\right)^{1/2} d \cdot \hat{E}$$

Here $\hat{E}$ is the electric field vector of the cavity, and $V_{eff}$ is the effective volume of the cavity mode, given by:

$$V_{eff} = |\boldsymbol{\alpha}_{\boldsymbol{\mu}}(\boldsymbol{r})|^{-2} = \int_V [\epsilon_r(r)|\boldsymbol{E}(\boldsymbol{r})|^2 d^3\boldsymbol{r} / \max(\epsilon_r(r)|\boldsymbol{E}(\boldsymbol{r})|^2)]$$

Here $\boldsymbol{\alpha}_{\boldsymbol{\mu}}(\boldsymbol{r})$ is the normalized mode function. Clearly, the larger the maximum electric field intensity of the cavity, the smaller the mode volume, and the higher the coupling strength.

In the weak coupling condition, the spontaneous emission (SE) rate of the emitter, $\gamma_{SE}$, coupled to a single-cavity mode can be understood from perturbation theory or classic models:[11,12]

$$\gamma_{SE} = \frac{8Q}{\hbar}\frac{|\boldsymbol{d} \cdot \boldsymbol{\alpha}(\boldsymbol{r})|^2}{4\epsilon_r\epsilon_0}\frac{(\gamma_{cav}/2)^2}{(\omega_0 - \omega_{cav})^2 + (\gamma_{cav}/2)^2}$$

Here, $Q$ is the quality factor related to the linewidth of the cavity mode $\gamma_{cav}$ by $\gamma_{cav} = 2\kappa\ (photon\ decay\ rate\ of\ the\ cavity\ field\ mode) = \omega_{cav}/Q$. Clearly, the SE rate is modified by the emitter-cavity interaction. Under resonance condition ($\omega_0 = \omega_{cav}$), we have the maximum emission rate $\gamma_{SE} = \frac{8\pi Q}{\hbar}\frac{d^2|\boldsymbol{\alpha}(\boldsymbol{r})|^2}{4\pi\epsilon_r\epsilon_0} = F\gamma_0$, where the intrinsic SE rate is $\gamma_0 = \frac{n_r d^2\omega^3}{3\pi\epsilon_0\hbar c^3}$, and $F$ is the Purcell factor describing the emission enhancement:

$$F = \frac{\gamma_{SE}}{\gamma_0} = \frac{3}{4\pi^2}Q(\frac{\lambda}{n_r})^3|\boldsymbol{\alpha}(\boldsymbol{r})|^2 = \frac{3}{4\pi^2}\frac{Q}{V_{eff}}(\frac{\lambda}{n_r})^3$$

Quantum mechanically, the emission can also be obtained by solving the master equation from the JC Hamiltonian.[13] Considering the coupling to an environment via decay, namely spontaneous decay of emitter into the vacuum modes from (rate $\gamma_0$) and photon decay of the cavity field mode (rate $2\kappa$), the SE spectrum in the weak coupling regime ($\gamma_0 \ll g \ll 2\kappa$) is described as:

$$S(\omega) = \frac{1}{2\pi}\frac{\gamma_0/2 + \frac{2g^2}{\kappa}}{(\omega - \omega_0)^2 + (\gamma_0/2 + \frac{2g^2}{\kappa})^2}$$

The emission rate in the cavity at resonance ($\omega = \omega_0$) has been enhanced from $\gamma_0$ ($g = 0$) to $\gamma_{SE} = \gamma_0(1 + \frac{4g^2}{\kappa\gamma_0}) \approx \frac{4g^2}{\kappa} \propto Q/V_{eff} \propto F$.[14] We again find that the interaction between the emitter and the cavity leads to enhanced SE rate. Particularly, the higher the quality factor of the cavity and the smaller the cavity mode volume, the larger the enhanced emission rate.

Unlike the weak coupling regime that the interaction between the emitter and cavity only modifies the spontaneous emission rate but not the emission frequency,[12] in the strong coupling regime the energy levels responsible for the emission are also altered. As easily understood from a harmonic oscillator model,[15] new hybrid modes different from those of the emitter or the optical system arise. Particularly, when $g > \left|\frac{2\kappa - \gamma_0}{4}\right|$, i.e. the coupling strength between the emitter and the cavity is stronger than the cavity and emitter decay rates, we would expect a peak splitting in the emission spectrum, with the energy difference being the Rabi vacuum energy,[13] given as:

$$\hbar\Omega = 2\hbar\sqrt{g^2 - \left(\frac{2\kappa - \gamma_0}{4}\right)^2}$$

In the true strong coupling regime, i.e. $g \gg (2\kappa - \gamma_0)$, we have $\hbar\Omega \approx 2\hbar g$. The expression for the SE spectrum is simplified as:

$$S(\omega) = \frac{1}{2\pi}\left(\frac{(\frac{\gamma_0}{4} + \frac{\kappa}{2})}{(\omega - \omega_0 - g)^2 + (\frac{\gamma_0}{4} + \frac{\kappa}{2})^2} + \frac{(\frac{\gamma_0}{4} + \frac{\kappa}{2})}{(\omega - \omega_0 + g)^2 + (\frac{\gamma_0}{4} + \frac{\kappa}{2})^2}\right)$$

As the coupling $g$ increases, the ideal emission spectrum evolves into two well separated Lorentzians of width (i.e. damping constant) $\Gamma = 1/2(\kappa + \gamma_0/2)$, centered at a frequency $\pm g$ away from the emission frequency $\omega_0$ (Figure S14B), respectively (i.e. a Rabi vacuum energy splitting of $\hbar\Omega \approx 2\hbar g$). Clearly, the increase of $g$ would split the new hybrid emitter-cavity modes further apart, favorable for achieving a large Rabi vacuum energy splitting for room-temperature quantum optical applications.

The observed phenomenon in Figure 3G is thought to be an implication of the presence of anti-crossing behavior. Improved coupling strength ($g$) between the emitter and the optical cavity could result in an anti-crossing manifestation in the energy levels (or spectrally in wavelengths) as a function of energy detuning (supplementary note 8 and Figure S14),[16] indicating that $g$ overcomes spontaneous emission of the emitter and the non-radiative energy loss from the cavity,[13] which is crucial to nanophotonics and quantum optics.[17,18] This anti-crossing phenomenon could be more easily understood from the classical oscillator models,[19] and further visualized in our simulation, where the AR of AuNRs is changed between 1.8 and 3.5 to modulate the cavity resonance mode (Figure S16). Clearly, a larger oscillator strength, which was modelled by larger extinction coefficients, could produce observable anti-crossing effect from simulation.

7. Fluorescent enhancement analysis

As mentioned in the 'METHODS' session in the main text, each sample in spectral analysis was prepared by adding a 20 μL drop of 100 pM AuNR-DO complexes at the center of a surface-modified fused silica chip. The overall intensity of fluorescent emission from various samples was quantified by an integral of the area under the measured emission spectra between 670 nm and 720 nm. Fluorescent signals from bare CES modified fused silica was taken as the background, whose intensity was subtracted from those of different samples before quantitative fluorescent enhancement was calculated. We assume that immobilized quantum emitters had the same surface density, considering that the initial concentrations of fluorescent species and surface immobilization methods were identical for all the samples.

8. Cavity mode volume calculation

Since our cavity structure consists of lossy material, the mode volume V is contributed from metal and dielectric material together. For the mode volume in dielectric medium, it is calculated following the traditional equation as shown in section 6, *i.e.*:

$$V = \frac{\int_V \varepsilon(r)|E(r)|^2 d^3r}{\max(\varepsilon(r)|E(r)|^2)},$$

where $\varepsilon(r)$ is the permittivity of the cavity material, $|E(r)|^2$ is the electric field intensity in position r inside the cavity.

For metallic material, $\varepsilon(r)$ in the integral is changed to $\mathrm{Re}(\varepsilon) + 2\mathrm{Im}(\varepsilon)\frac{\omega}{\gamma}$, where $\omega$ is the incident light angular frequency and $\gamma$ ($7.14\times10^{13}$ $s^{-1}$) is the damping rate based on Drude model.[20-23]

Since the highly enhanced electric field is confined within 4-5 nm inside the nanogap between the DNA origami assembled gold nanorods, and the electric field intensity greatly drops outside the nanogap, $\int_V \varepsilon(r)|E(r)|^2 d^3r$ is integrated in a 8 by 8 by 8 nm cube centered at the nanogap.

The calculated Purcell factor $F = \frac{3}{4\pi^2}\frac{Q}{V_{eff}}(\frac{\lambda}{n_r})^3$ for a AuNR dimer of AR=2.8 (33 nm long, 12 nm wide) with 4 nm gap is $2.42\times10^6$, which well matches the experimentally derived value.

9. Lifetime analysis

The lifetime measurement was conducted with a PPD detector and the raw data was fitted using DAS6 software from Horiba. This fitting assumed that the output signal was the convolution of an exponential decay function with a Gaussian response. To rule out the influence of instrumental response, a data set following prompt was also collected by measuring the fluorescent signal decay using bare CES modified fused silica. After subtracting the scattered photons from the non-fluorescent media, the exponential decay function could be deducted by fitting the data with the model below:

$$y = y_0 + \int_{-\infty}^{+\infty} Ae^{-tx} \otimes Response\ \ dx$$

Here $y_0$ and $A$ are two fitting parameters, and $t$=1/$\tau$.

The fitting results of lifetime measurement show minimal residue for all the samples (Figure S19). Besides, the $\chi^2$ method was employed to judge the quality of the fit, following previous literature.[24] The $\chi^2$ value is calculated to be 1.3, 1.4, 0.3 for free Cy5, DR and AuNR-DR, respectively, indicating decent fitting goodness.

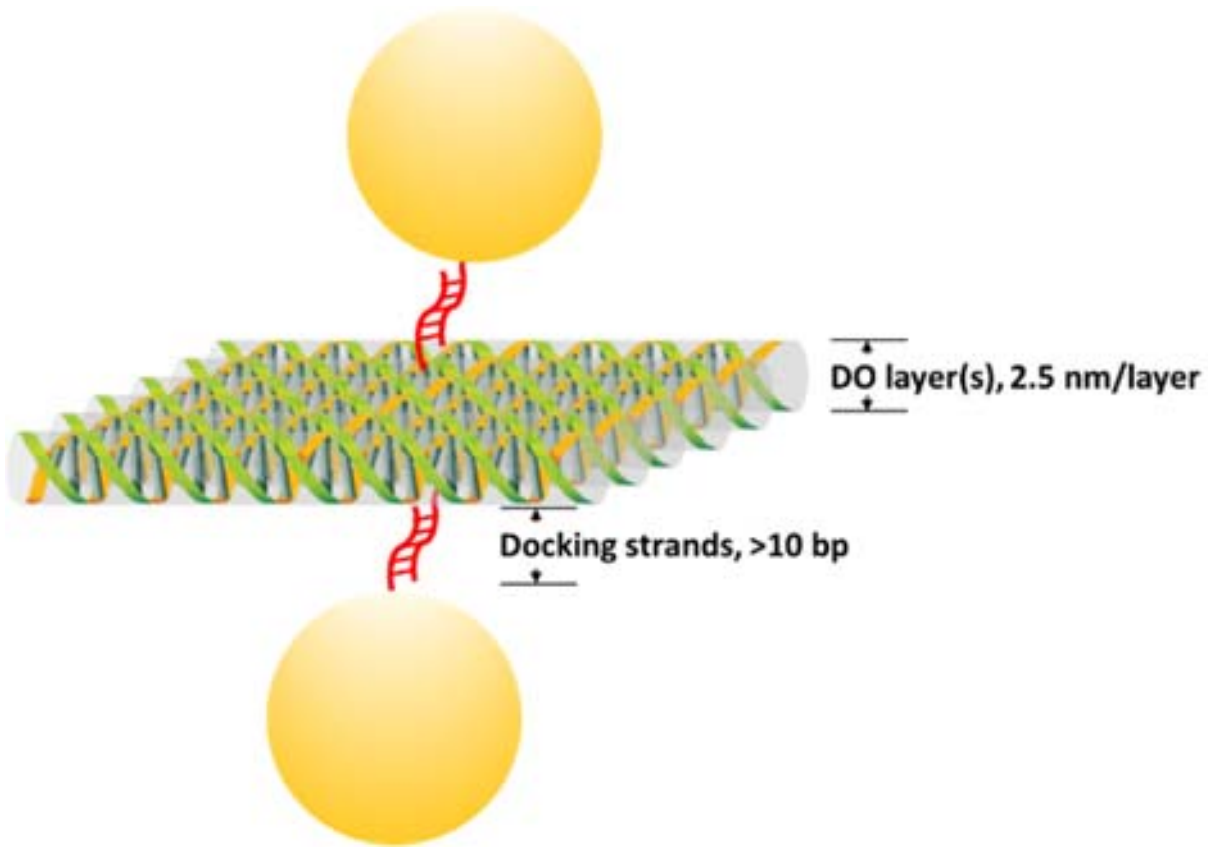

Figure S1. The nanocavity size created by the conventional DNA assembly strategies has a large minimal distance, which at least equals to the sum of the lengths of the two docking strands that hybridize with the capping strands on the AuNPs (including the lengths of the linker molecules) and the thickness of the DO.

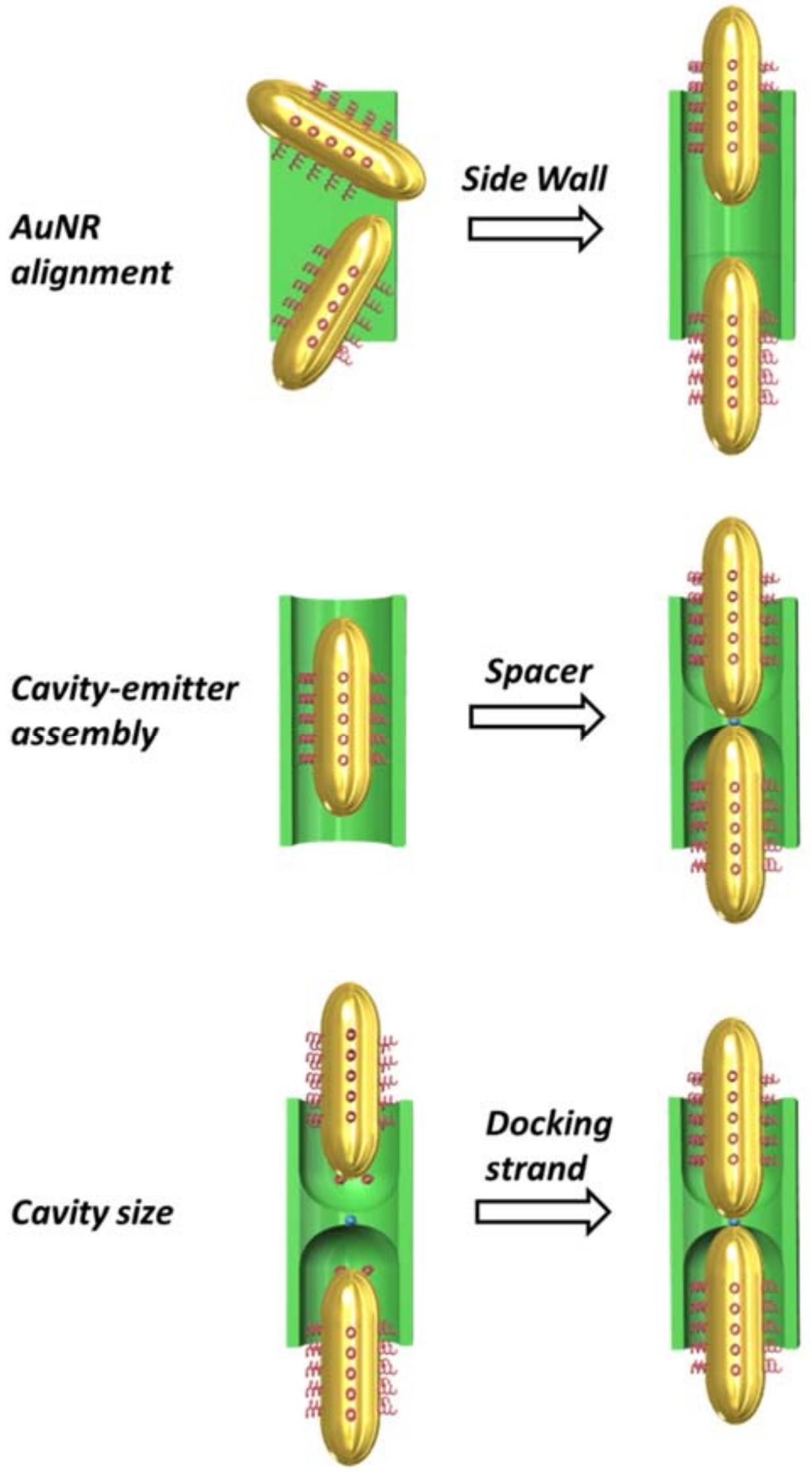

Figure S2. Design principle of DR. Docking strands hybridization only occur on the same groove side of the DR, separated by the spacer in between. Side walls of the grooves limit the orientation freedom of the AuNRs, and help to align them into linear dimer with a linear conformation. The spacer in the middle controls the gap separation of the dimer and allows the precise attachment of the emitter in the center position of the cavity once the two AuNRs snuggly are attached inside the grooves. The uneven distribution of the docking strands with higher density close to the spacer minimizes the nanocavity gap.

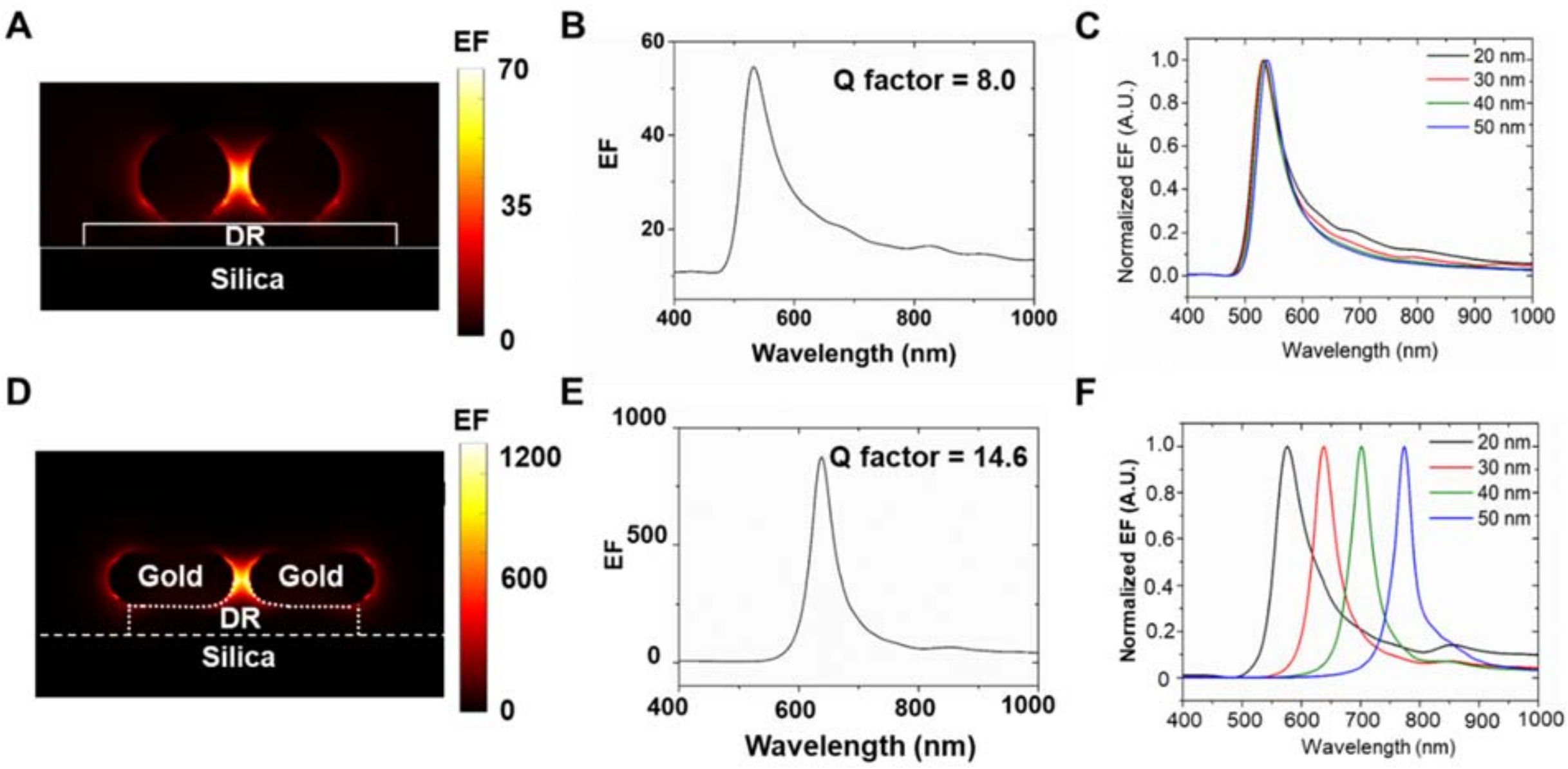

Figure S3. Simulated properties of plasmonic nanocavities formed by AuNSs and AuNRs. (A) EF distribution of a coupled AuNS (18 nm in diameter) dimer with 4 nm gap. (B) Wavelength dependence of calculated EF at the center of the nanocavity in (A). (C) Dependence of plasmonic resonance wavelengths on the diameter of AuNS in the dimer. (D) EF distribution of a coupled AuNR (12 nm width, 30 nm length) dimer with 4 nm gap. (E) Wavelength dependence of calculated EF at the center of the nanocavity in (D). (F) Dependence of plasmonic resonance wavelengths on the length of AuNRs in the dimer.

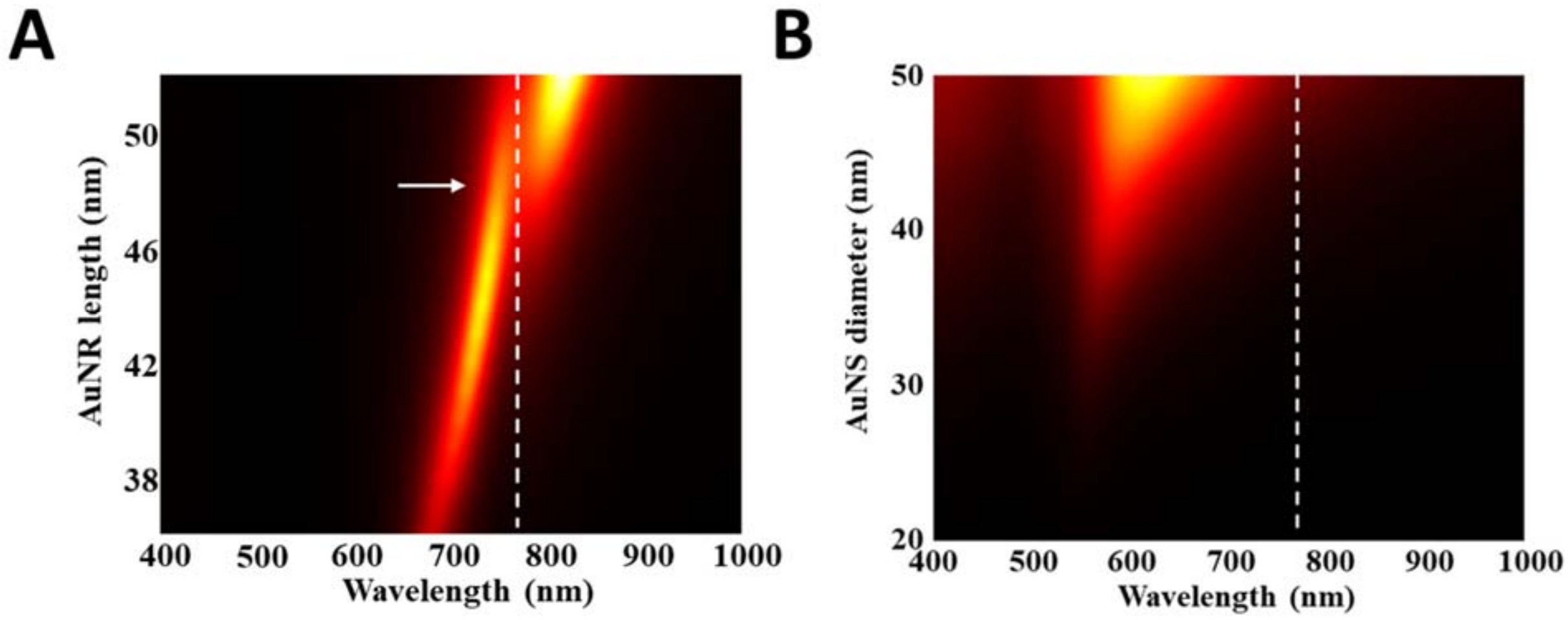

Figure S4. Demonstration of the versatility of plasmonic nanocavities formed by AuNRs. (A) Anti-crossing effect calculations for AuNR based plasmonic nanocavity. The width of AuNRs and the gap size is fixed at 12 nm and 4 nm, respectively. (B) Anti-crossing effect calculations for AuNS dimer based plasmonic nanocavity. The gap size is 4 nm. The white dashed lines in (A) and (B) indicate the emission peak wavelength of a Cy7 dye.

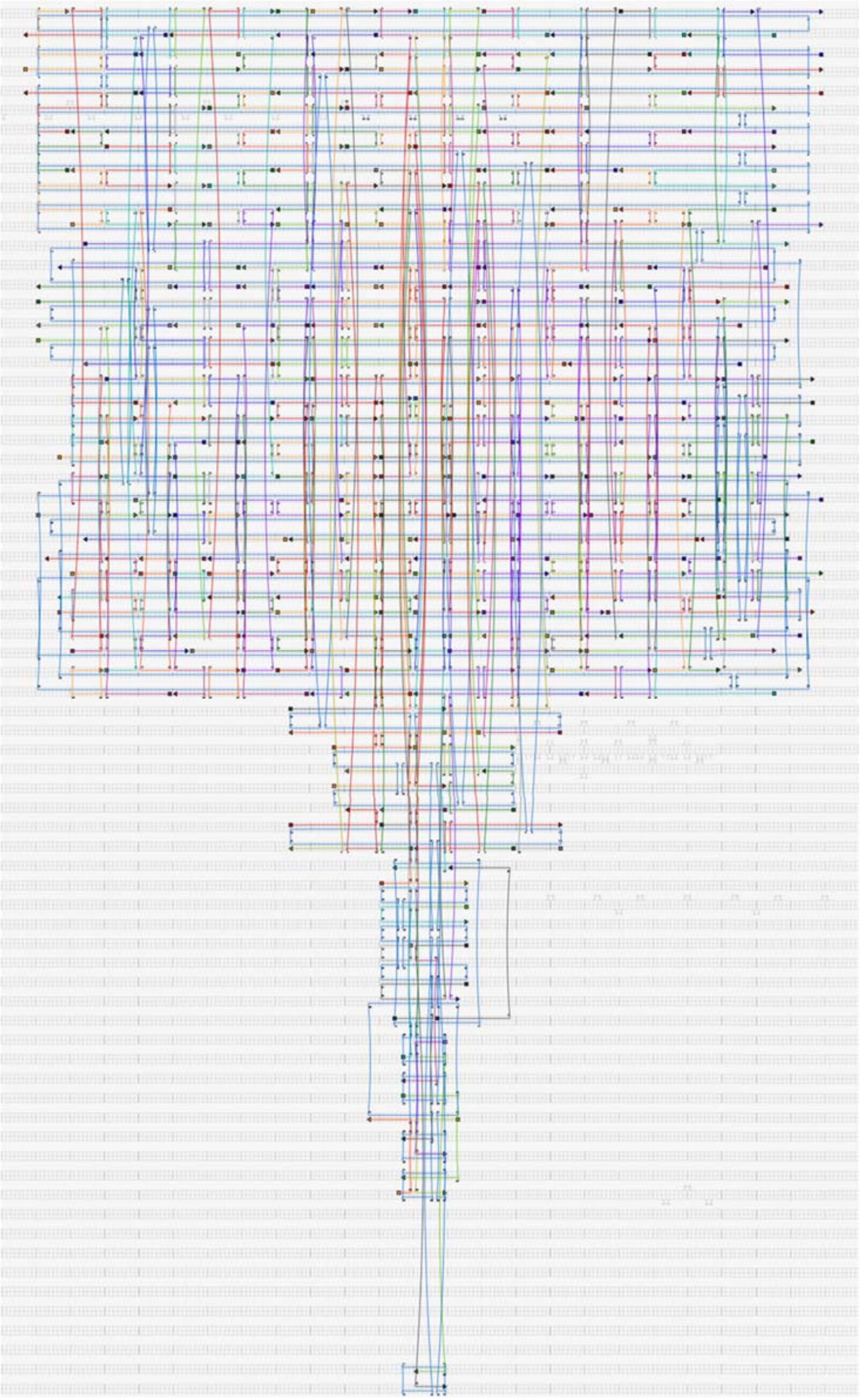

Figure S5. The design of DR by Cadnano 2.0.

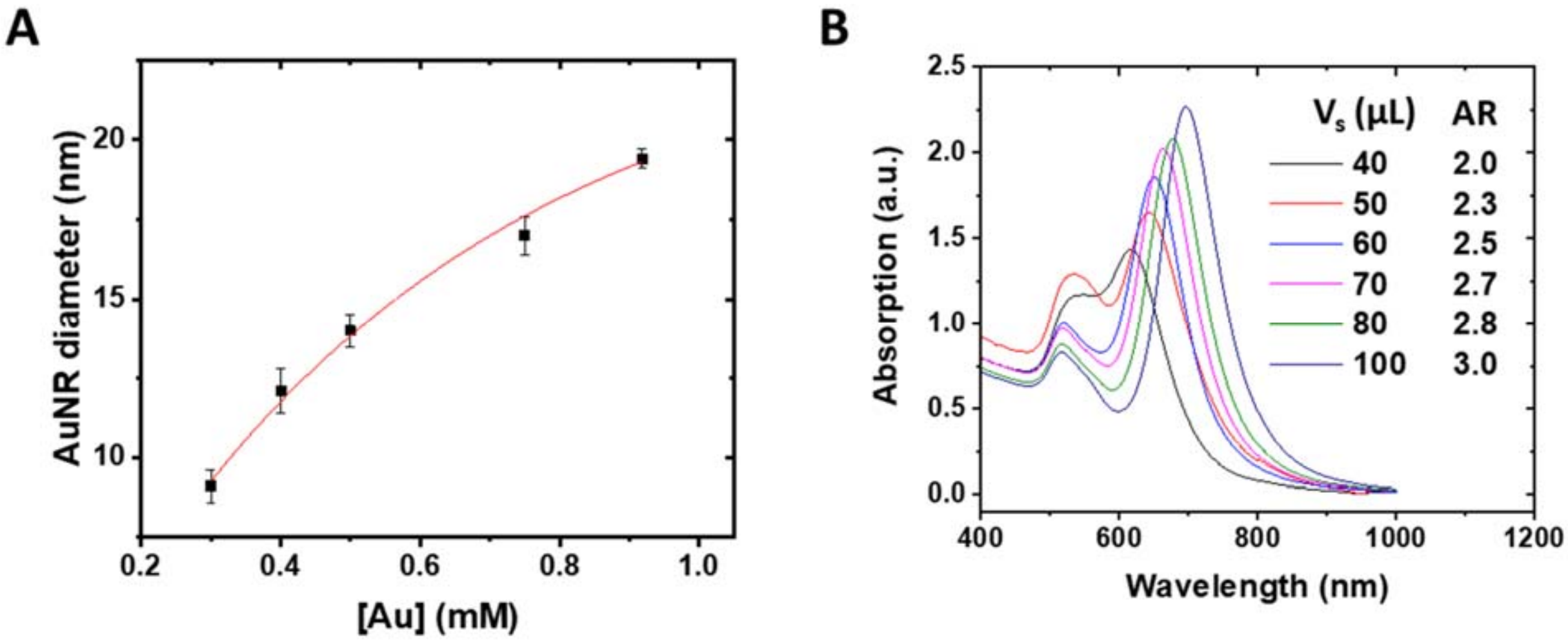

Figure S6. Tuning geometric parameters of AuNRs by adjusting experimental conditions of the synthesis. (A) The width of synthesized AuNRs as a function of Au(III) ion concentration. (B) SPR absorption red shifted and intensified as the volume of $AgNO_3$ solution increased. The corresponding ARs of the AuNRs were calculated from size measurements based on TEM images. The Au(III) concentration was fixed at 0.4 mM.

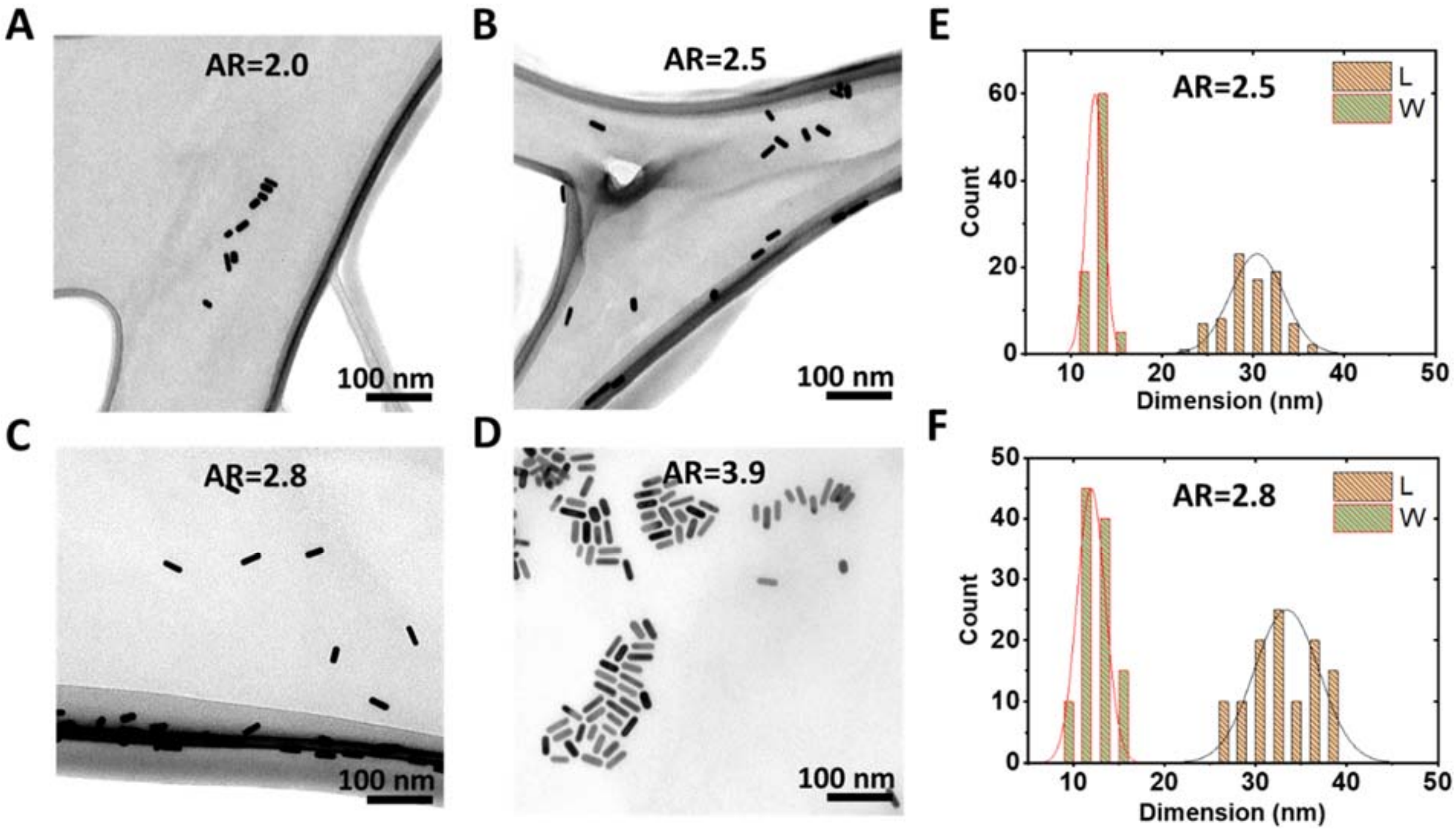

Figure S7. Morphological characterization of the synthesized AuNRs. (A)-(D) The successful preparation of AuNRs with various AR confirmed by TEM. (E)-(F) Representative size distributions of AR=2.5 and AR=2.8 AuNRs, measured from the TEM images.

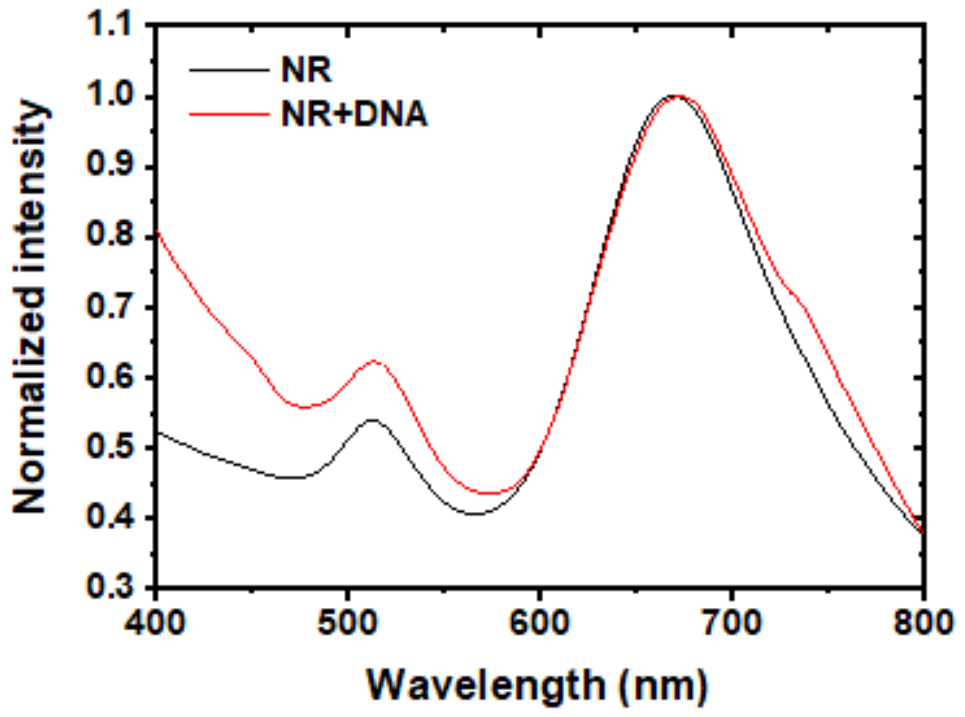

Figure S8. UV-Vis spectra of AuNRs before and after capping with thiolated DNA.

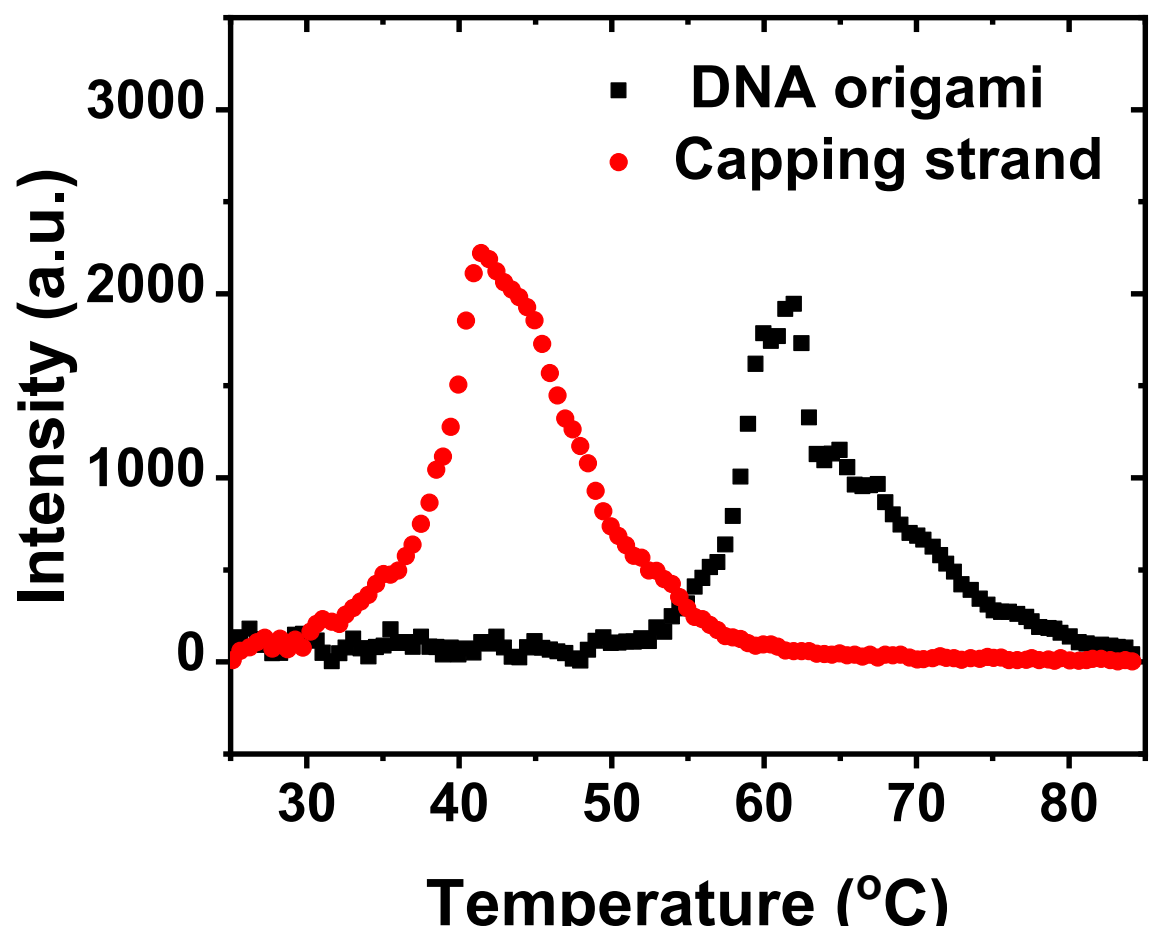

Figure S9. Measured melting temperature of DNA origami, and the capping strand binding with the complementary strand. A significant difference in the two melting temperatures allows the two step annealing for the assembly process.

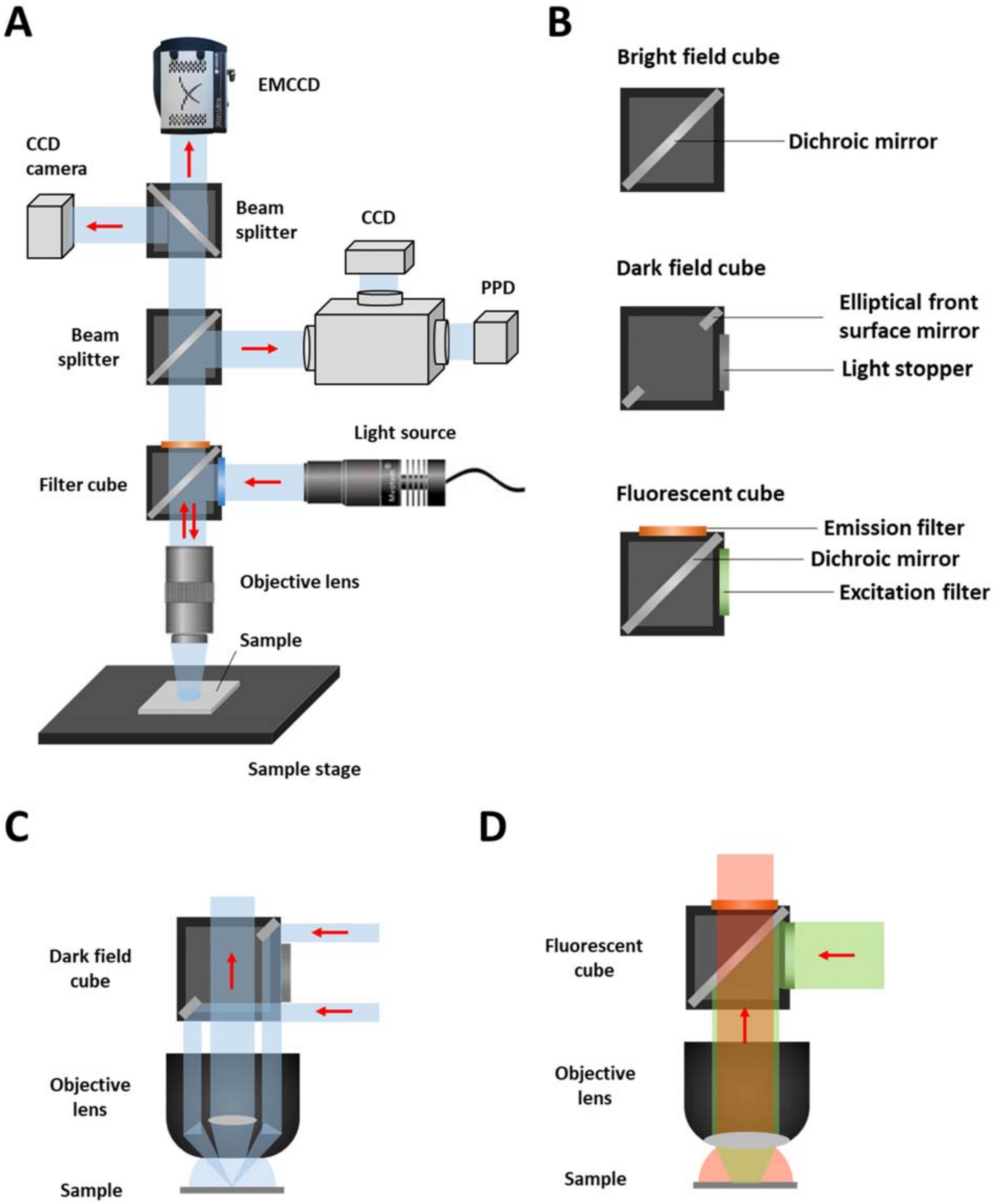

Figure S10. Scheme of the setup of the microscope-coupled UV-Vis spectrometer. (A) Scheme of the overall setup. Red arrows indicate the flow of light beams. (B) Detail scheme of the filter cubes. (C) Measurement principle of dark field mode. (D) Measurement principle of fluorescent and lifetime mode.

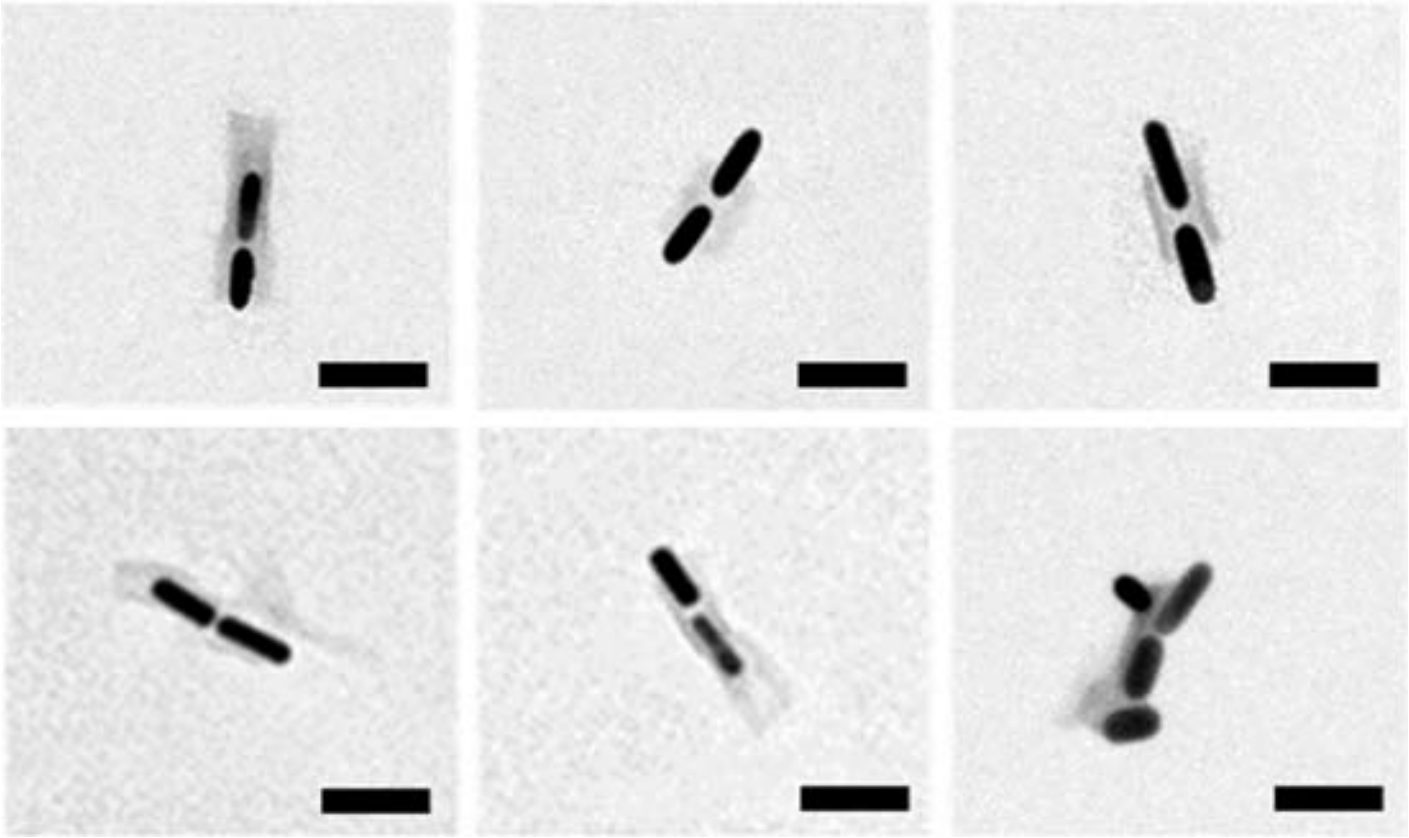

Figure S11. Additional TEM images of AuNR dimers formed through DR guided self-assembly. Scale bars: 50 nm.

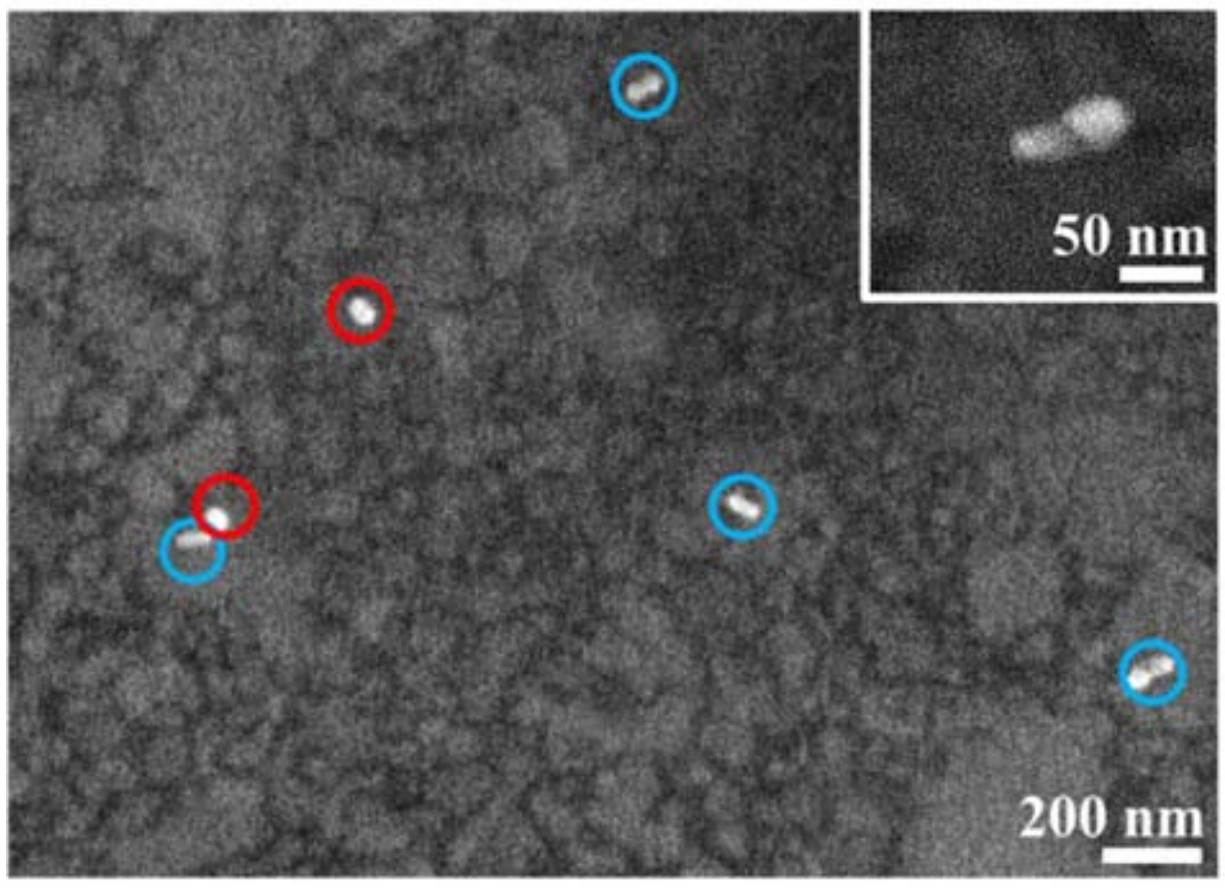

Figure S12. SEM image of surface immobilized AuNR-DR complexes. AuNR dimers and monomers are indicated by blue and red circles, respectively.

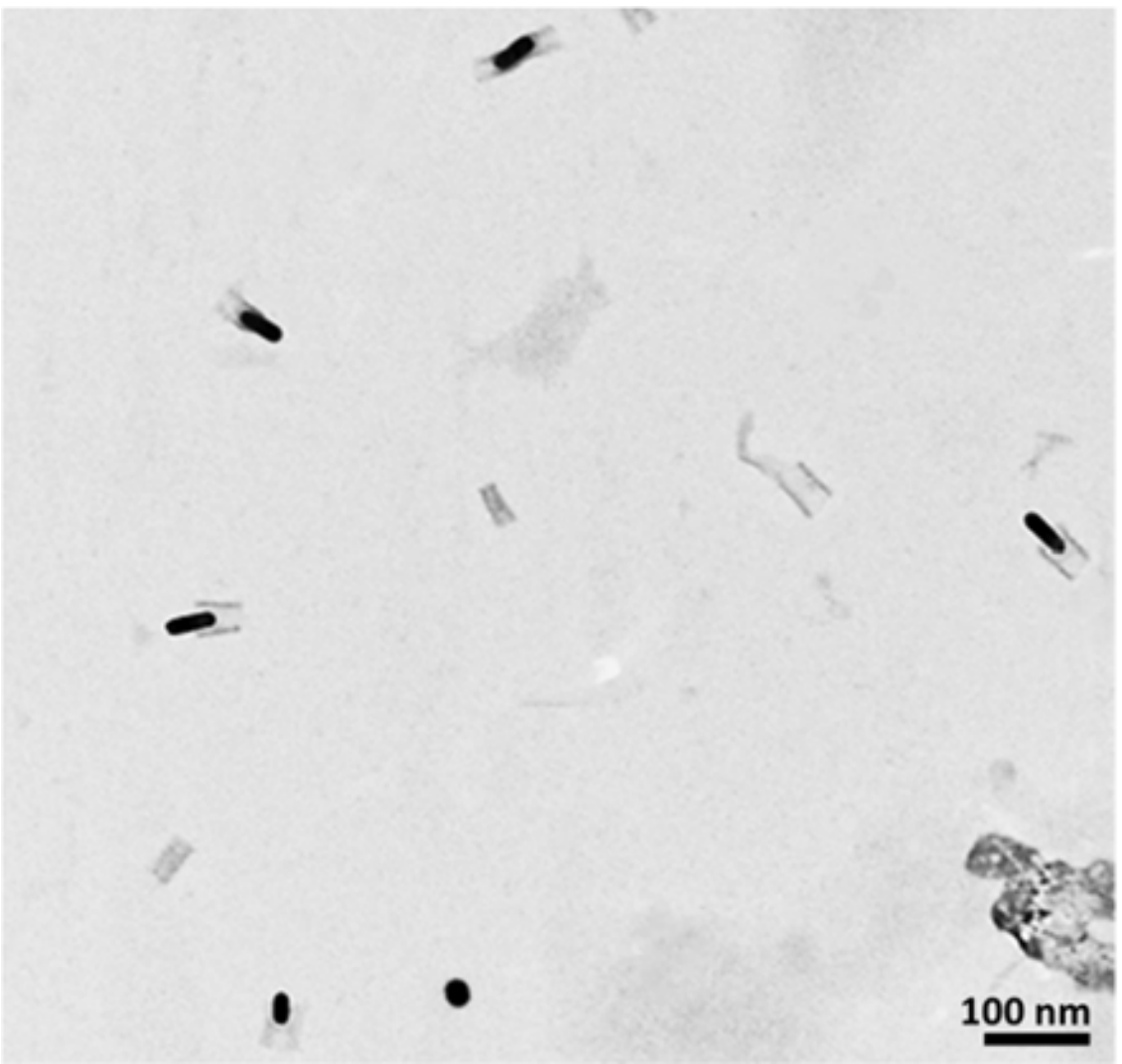

Figure S13. TEM of AuNR monomers attached to DR.

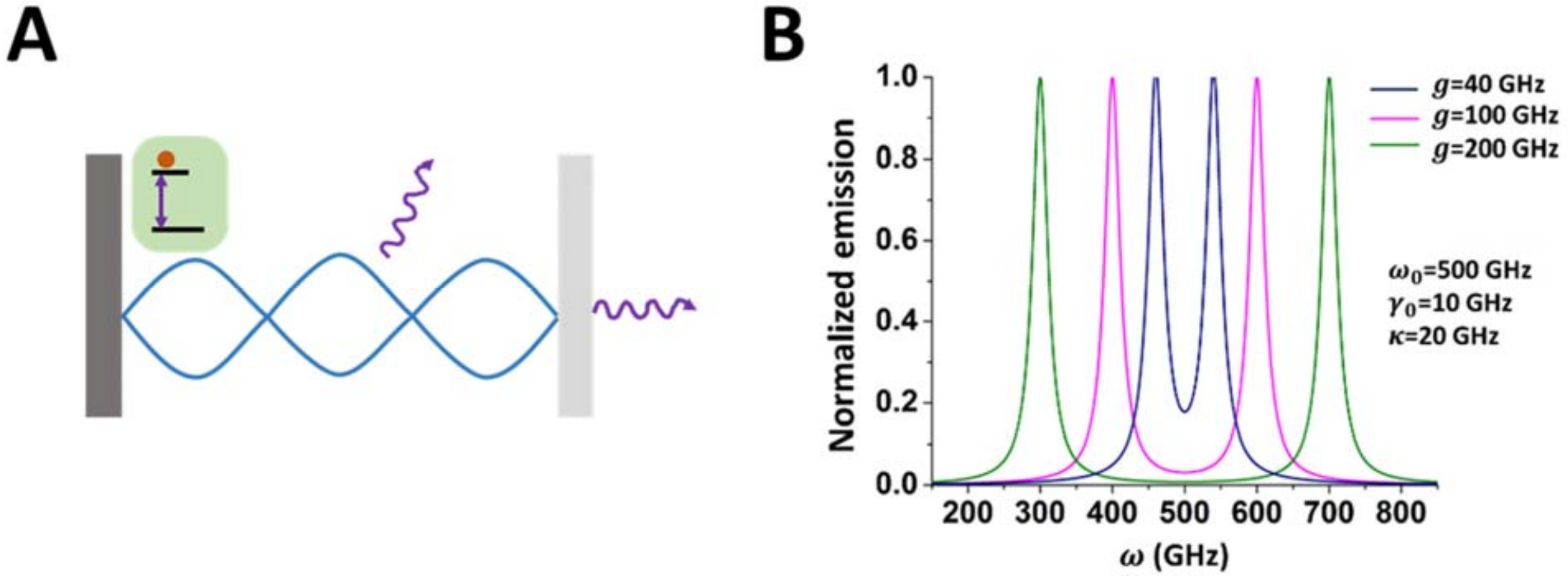

Figure S14. Understanding of light-matter interaction. (A) Schematic of a two-level emitter in a cavity. (B) Calculated spontaneous emission spectrum at different coupling strength $g$.

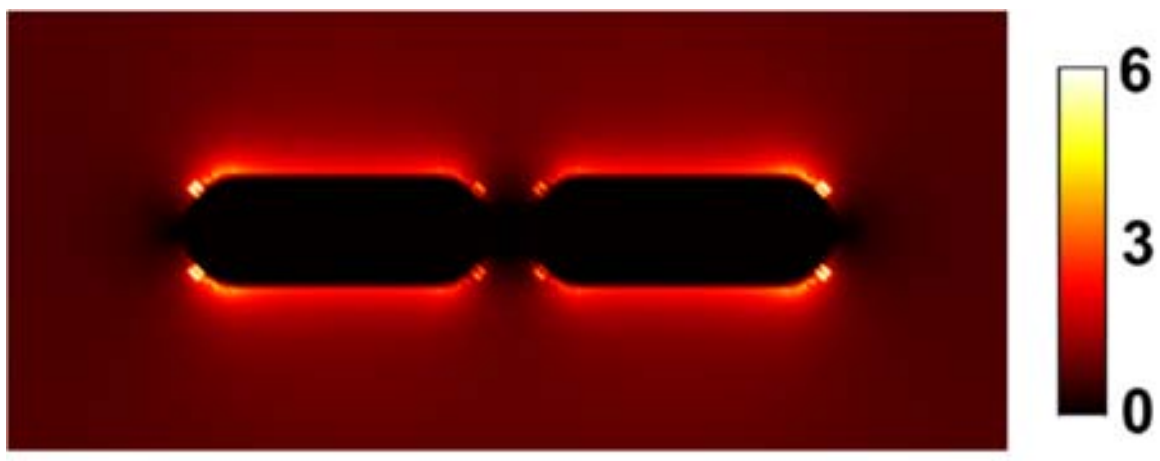

Figure S15. Distribution of EF in the transverse mode of an AuNR (33 nm long, 12 nm wide, 4 nm gap) dimer . The incident wavelength is 669 nm.

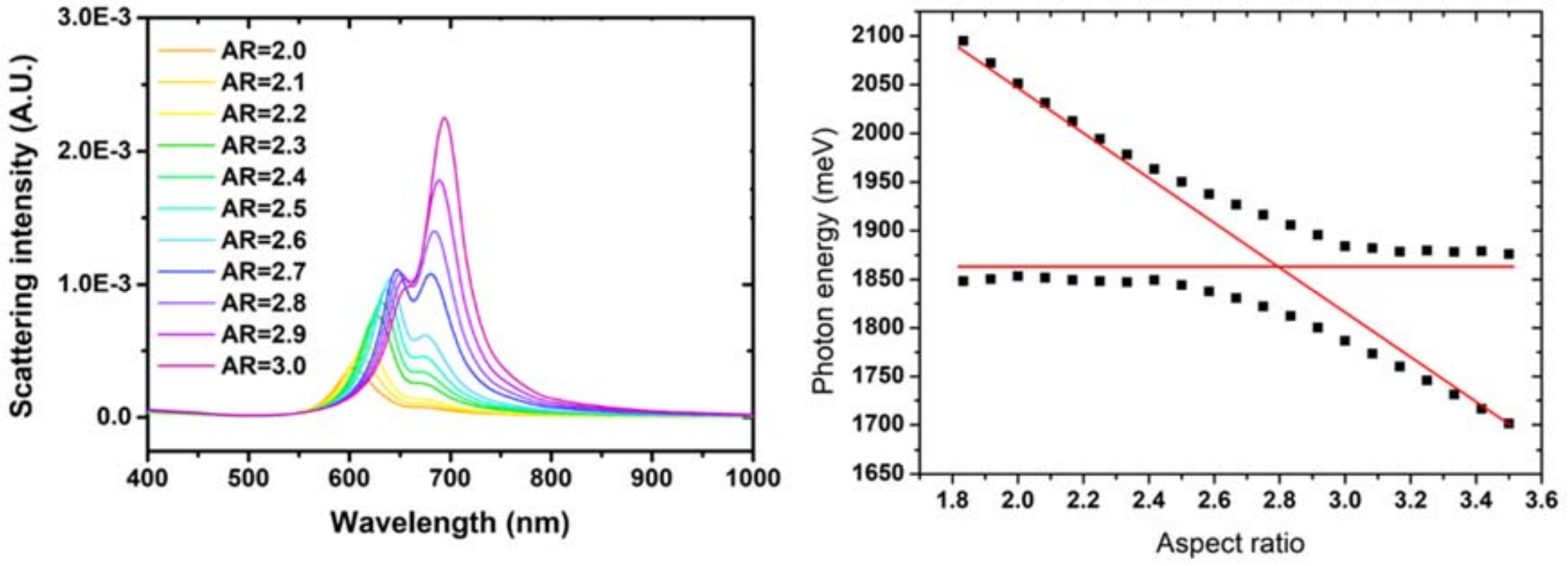

Figure S16. Simulated results showing the anti-crossing effect as the AR of AuNRs varied (width fixed at 12 nm). Left: Scattering spectra. Right: Peak positions (in meV) as a function of AR.

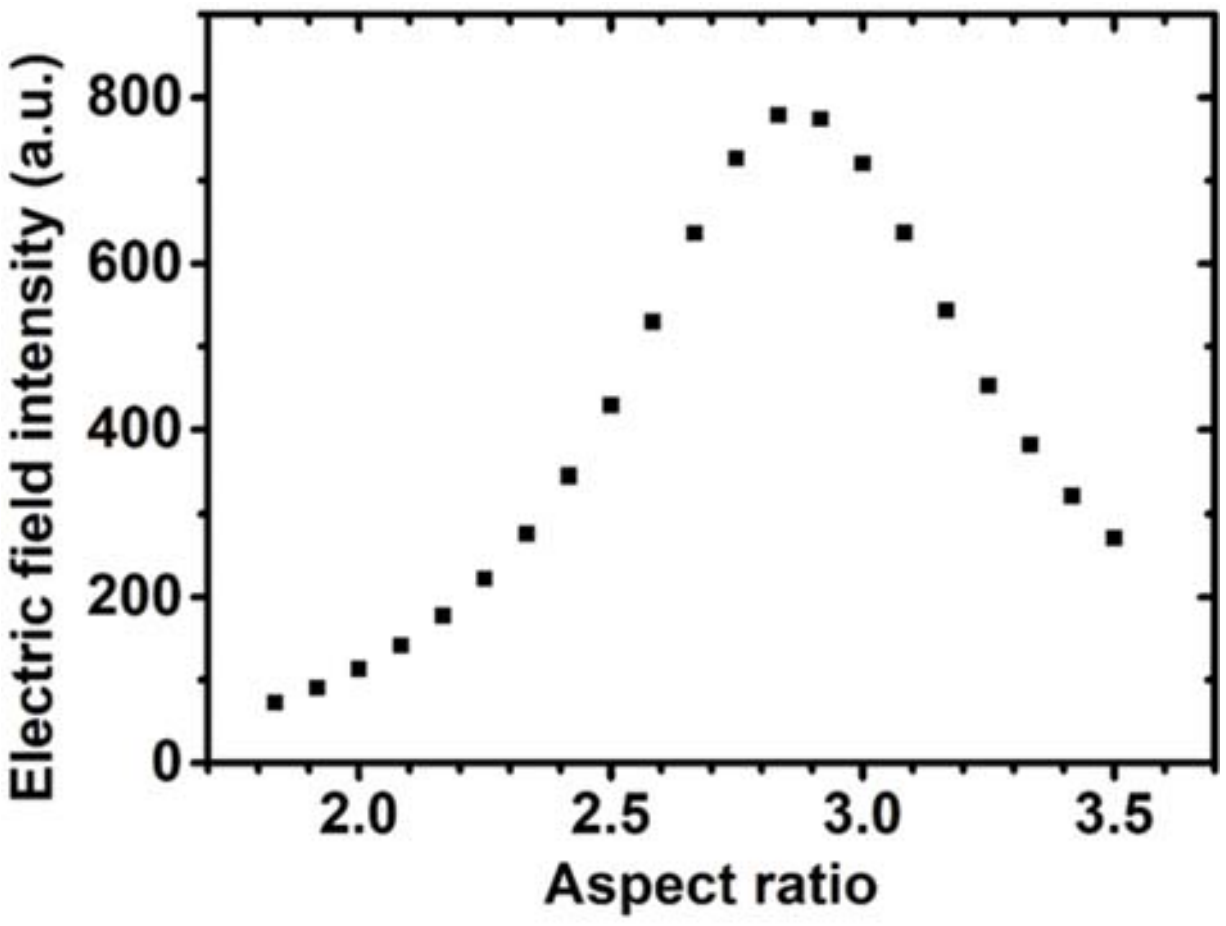

Figure S17. Calculated electric field intensity at the nanocavity as a function of the AR of AuNRs.

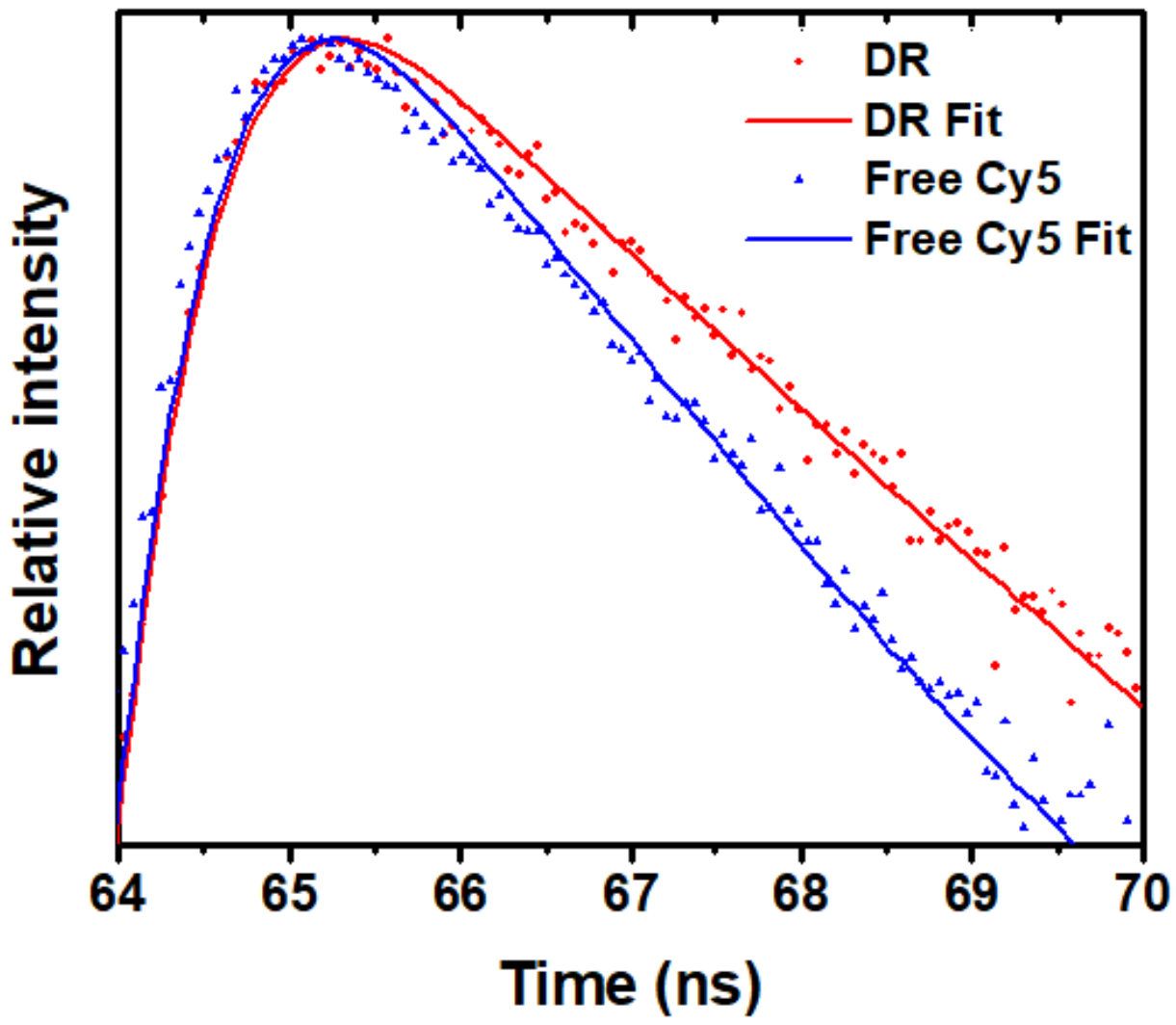

Figure S18. Fluorescent lifetime measurements of Cy5 in the solution and on the DR.

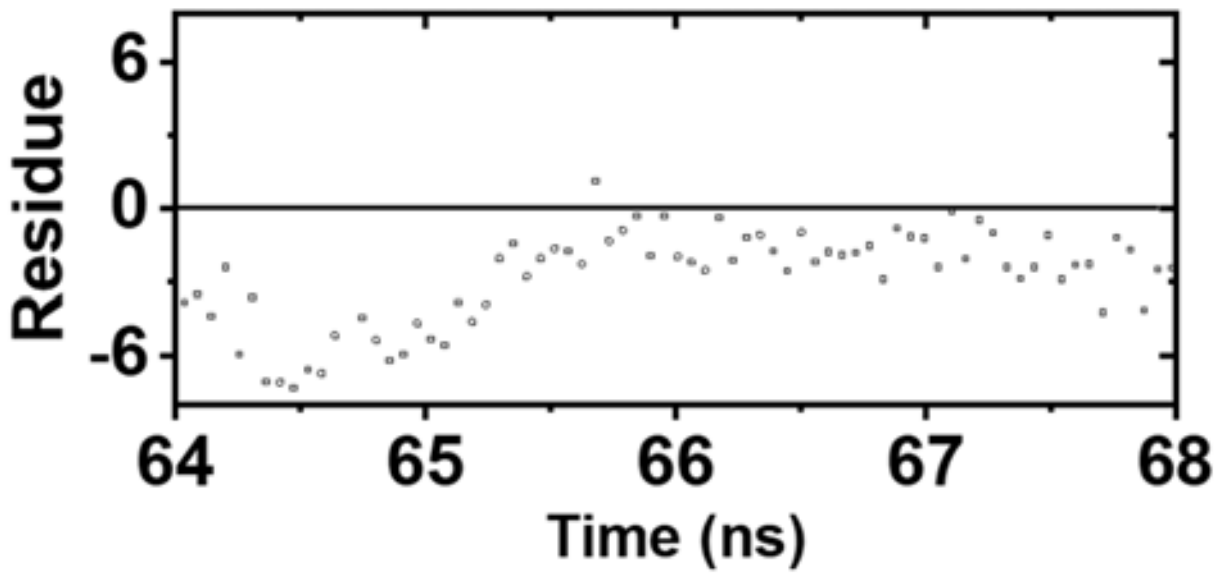

Figure S19. Fitting residue of fluorescent lifetime measurement of free Cy5.

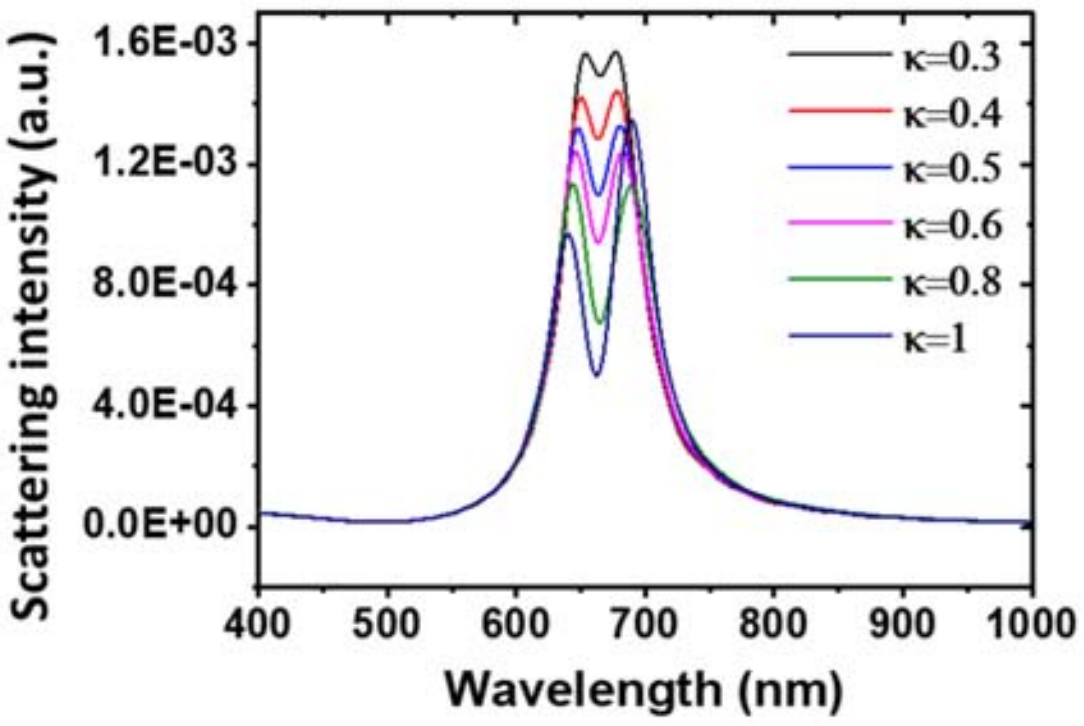

Figure S20. Simulated effect of κ on the scattering spectra. The curves are named after the maximum κ value in each case.

Table S1. Representative studies of emitter interaction with nanocavities formed by plasmonic nanoparticles*

| Design scheme | 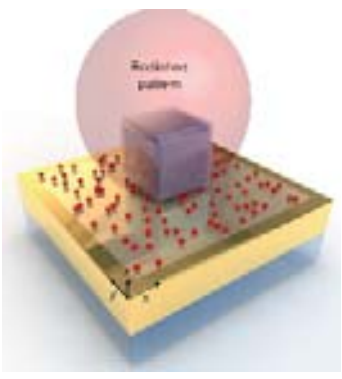 | 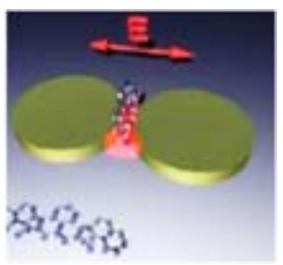 | 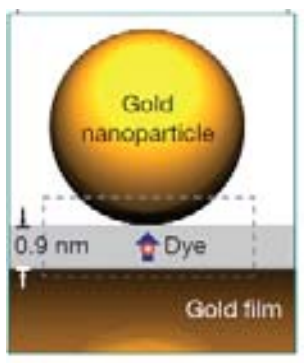 | 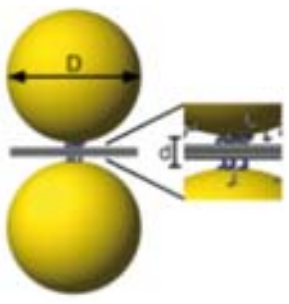 | 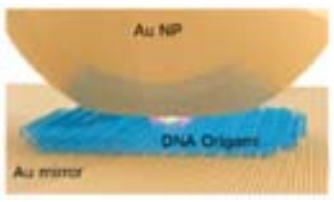 | 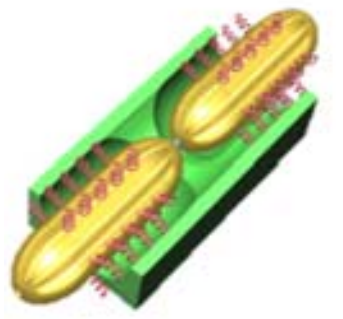 |
|---|---|---|---|---|---|---|
| Cavity design | Nanocube on metal mirror[25] | Nanodisk dimer[26] | Nanosphere on metal mirror[27] | Nanosphere dimer[1] | Nanosphere on metal mirror[28] | Anisotropic nanocavity (This work) |
| Critical structural dimensions | ~75 nm nanocube, ~10 nm gap | 60-115 nm disk, 15 nm gap | 40-60 nm sphere, 0.9 nm gap | 30-60 nm particle, 5-10 nm gap | 80 nm particle, 5 nm gap | 12 nm wide AuNRs, AR=2-4, 2-3 nm gap |
| Near-field intensity enhanceme nt | ~200 | ~260 | ~400 | NA | ~3.5×10$^3$ | ~4×10$^3$ |
| Cavity volume | ~6×10$^4$ nm$^3$ | ~5×10$^3$ nm$^3$ | ~40 nm$^3$ | ~500 nm$^3$ | ~10$^3$ nm$^3$ | ~20 nm$^3$ |
| Calculated cavity Q factor | ~10 | ~8 | ~16 | ~8 | ~8 | ~15 |
| Calculated $Q/\sqrt{V}$ | 0.04 nm$^{-3/2}$ | 0.1 nm$^{-3/2}$ | 2.5 nm$^{-3/2}$ | 0.4 nm$^{-3/2}$ | 0.3 nm$^{-3/2}$ | 3.3 nm$^{-3/2}$ |
| Emitter number control | Statistical, adjust emitter concentration | Very challenging | Statistical, adjust emitter concentration | Very challenging | Deterministic | Deterministic |
| Emitter placement accuracy | Random | Random | Random | Random | Nanometer accuracy | Nanometer accuracy |
| Purcell factor | ~5×10$^2$ | ~5×10$^3$ | ~3.3×10$^6$ | ~4.7×10$^4$ | ~4×10$^3$ | ~2.4×10$^6$ |
| Emission frequency tuning | Difficult | Narrow range, adjust particle size | Narrow range, adjust particle size | Narrow range, adjust particle size | Narrow range, adjust particle size | Wide range, adjust the AR of AuNRs |

* Green color indicates desired properties.

Table S2. List of all the DNA sequences

| Group | Name | Sequence (5'→3') |
|---|---|---|
| | S1 | TTCTGGCCCCTTGCTGGAACGGTATCACGCTG |
| | S2 | AATGAAATACGTCAAACACCCAGCTACAATTTTATCCTGACTC ATCGA |
| | S3 | GCTGAGAGGGGAGCTATGCAACAG |
| | S4 | CACAGACAATATTTTTGAAGTAGCAATTGTCCATC |
| | S5 | CCGATATAAGAATCCTAATTGAGA |
| | S6 | ACGCCTGTGGCGACATTCAACCGACGCATTAG |
| | S7 | TCTTTCCTTATCAGCCTAATTTGCCATTT |
| | S8 | GCATCACCGTATAACGCATTTTGATTT |
| | S9 | CAGGAGGAGAATCAAGCCGCCGCCCAGAATGGAAA |
| | S10 | CATATGCGTTATACAACAAGAAAAATACCGAT |
| | S11 | TTAGTAAATTTCAACACAGCAGCG |
| | S12 | TTGCCATCTTTTCATAATCATTAGCGG |
| | S13 | ATATTAGAAATGTAAAAAAACGAA |
| | S14 | TTTTTAATGCCCCCTGCCTATTTCCCCTCATAAGACTCCT |
| | S15 | TTTTAAAACAATTAATGCGCGAACTGAGGATTTAGAAGTATTA |
| | S16 | TGAGAATAGAAAAATTCATAAGCGCTAA |
| | S17 | ACGTTAATACCAAAAAGAAGTTT |
| | S18 | ATGCTGATATTGAATCCCCCT |
| | S19 | ACGGGTATCCATCCTAATTTACGAAAAGCCTG |
| Structural strands | S20 | TTTGAACTGGCATGATAAGTTACCTTT |
| | S21 | CAGGGAAGTTGAGGGAGTCACAATTCAGTACCAGGCGGAT |
| | S22 | ATTTCAACTTTAATCATACTAATAGTAGTAGCAAGAATTAGCA AAATT |
| | S23 | CCAGCAGCTTTACATTGGCAGATT |
| | S24 | AACGATTTACTTGAGCATACATACCACCGTACT |
| | S25 | CCGAAGCCCGTAGAAACATTTGGGAGAGCCAC |
| | S26 | AGAGTCAATAGTTATTATTATTATTTGAATTTA |
| | S27 | AAAGAGTCACTTCTTTCAAAAGAAAGCTAACTACGAGCCG |
| | S28 | TTTAAACAATTCGAGTTTGAGTGCCCGAACCAAAGAAACCACC TTT |
| | S29 | ACCGCGCCCAATAGCAGAGGTTTTCATAAAAA |
| | S30 | GGATTCTCGATAGGTCACGTTGGTAGGGCGCG |
| | S31 | TTTTATTAAGAGGCTGGTTAGCGTAACGATTT |
| | S32 | TTCAGGCTATATTTTGGGTAAAGA |
| | S33 | CATCTGCCCACCACACACGCTGGTCAGAACAATATTACCG |
| | S34 | GCCCAATAGGAACCCAAGCCACCATCATCGGC |
| | S35 | TTTGAGAAACAATAACGGATTCGCGGCGTTAAATAAGAATAA ACATTT |
| | S36 | CCACCAGCAGTGTTTTGAGTAGAATGGTTCCGGCCCGCTTAAT TGTTA |
| | S37 | TAATAAGAATGGAAACAGTACATATGAATTAC |

| | |
|---|---|
| S38 | TAACCGTTTGGCTATTAGTCTCTCGTATTAAATCCTTT |
| S39 | ACATTATTACAGGTAGGACATTCATTAAGGACGTT |
| S40 | ACAGGATTCCTTTATTTCAACGCAGGTGAGAA |
| S41 | GATAATCACCGTTCTAGCTGATAACGGTTGTAGTGTCTGGAAG TT |
| S42 | AGTATAGGCGCGTTTCCCTCAGAGTTCCAGTAAGCGTCA |
| S43 | ACAGGAGGCAGAACCGCGCCACCC |
| S44 | CAATTCCAGTTGGGAAAAAATCCCACGTGGACGTGAGGC |
| S45 | CAAGAAACATAAAAGAAAAGGTGA |
| S46 | TGAGAGAGAGAGGCGGTCTAGAGG |
| S47 | GCAGTATGGCTTGCTTTCGAGGCAATAATAACGGAAT |
| S48 | CTTATTACTCAGAACCGAAACGTCACCAA |
| S49 | AGGAAGGTGTGAGGCGGTCAGTATTTAGACAG |
| S50 | CCTGGCCCGAAAAACGCTCATGGAATGGATTAAAATGAAAAT CAAACC |
| S51 | TCCTGAGAAGAAGATATAGAACCC |
| S52 | ATAATATCTAAACCAAGTACCGCAATCTTACCAAATAAGA |
| S53 | TACCTTATTGAATAAGGCTTGCCCACAAAGTA |
| S54 | CACCGTAATCAGTAGCGACTTTAGTACCCACCCTC |
| S55 | AATACATTTGATAGCCCTAAAACA |
| S56 | CAGTAACAGGACAGATGAACG |
| S57 | ATATGTAAGAGAGACTACCTTTTTACAACAAC |
| S58 | ATGCAACTTGGTCATAATATTCAA |
| S59 | AAAGACAGCTACGAAGAAACGGGTAAAATACGTTT |
| S60 | GTCGTGCCGCAGGCGACCAGCTGGGGGACGACGGCCTTCC |
| S61 | CTTAATGCGCCGCTACGTAGATGGTGGGTAAC |
| S62 | GCGAGAAAATCGCCATATTTAACA |
| S63 | GACGACGAACGCCAACATGTAATTTGAATAACCTTGCTTCAAA CAAAA |
| S64 | CCGCTTTTAAAGTAATAAGGCTTATCCGGTATT |
| S65 | TCAATATGGCTGTTTCCTATTACG |
| S66 | TTTTGAAACCATCGATGGTCAGACGATTGTTT |
| S67 | CTCAACAGCGCCTGTTCATCGCCC |
| S68 | GTAAATTGCTTTGACCCCCAGCGAACACTAAAACACTCATTTT |
| S69 | CGTAATCAAAAGTACGCCAAAAAC |
| S70 | CTATCTTAAGTTGCGCGCCTTTAA |
| S71 | ACGCATAATGAAAATCTCAAAATCTGAAA |
| S72 | GATTATACTTCTGAATATTAACCAGGC |
| S73 | ATTATCACCGTCACCGTTTGTTTAAGCAATAG |
| S74 | GCAAACAACAGGTCATTGCCTGAGTCATACAGGCAAGGCAAT TAACAT |
| S75 | CTTCACCGTACTATGGTCGTTAGAATCAGAGC |
| S76 | ATTCACAAACAAATAAATCCTCATTAAAGCAGCATTG |
| S77 | ACATATGAGAGACGCGCG |
| S78 | TAGCCGGATCATTTCAATTACCTGTTTAGTTA |
| S79 | GATTAAGTGCGCATCGAATGTGAG |
| S80 | TCCCAATTAACCTGTTATTATGACCCTGTAAT |
| S81 | CGAGTAACAACCCGTCTAATCGTA |
| S82 | GAGGCTTTGGGGTAATAGTAAGCTCAACATGTTTAGAGGAAACCT |
| S83 | GAAAAGCCCCAAAAACTCGCGTCTGACAGTATCGGCCTCA |
| S84 | GCAGAACGTAGGGCTTATAGGTCT |
| S85 | CACCGAGTATCGCCATTCGTGG |
| S86 | TTTCGCTCAATCGTCTGAAAATACCTATGCTTTCC |
| S87 | CGCGTAACAGTTTGAGCGAAAGGG |
| S88 | AAGCCTCAAAGGTGGCATCAATTCTTGTGAAT |
| S89 | AAAATAATAGGAAGATCCTCTTCGCTGTGTGATCCAGTCGGGT |
| S90 | GAAAAACCTTTGATGGGAACTCAAACTATCGG |
| S91 | CACCAGTCCCAGCCATAACAGGAG |
| S92 | TTTGTTACAAAATAAACAGCCATAATCACCAG |
| S93 | TCCAACGTCAAAGGGCGGAAGATC |
| S94 | AAGATGATATCCGCGACCTGCTCCAACCGGATCATAAATC |
| S95 | AGGTGTATATAAAGGTCAAAAGGA |
| S96 | TATCAACAAAGCCGTTTTTATTTTCTATTTTGAATGAAAA |
| S97 | CCCTGACTATTATTCTAAAAATCAGGTCTTTACAAAAGGA |
| S98 | TAGCTATATTTTCATTGCTCATTA |
| S99 | AGCAGATAAACGTGAATTTCTTAAACAGCTTG |
| S100 | TTTATTTTGGGAAGGTGTTTCGTC |
| S101 | TTTTTATCAGAGAACCTACCATATTTT |
| S102 | CTTTTTTATTATACCAAGCGCGAATGACGAGAAACACCAG |
| S103 | CCCTTATTCACCCTCAATACAGGAGTGTACTG |
| S104 | CCGGAAGCAGATTAAGAGGAAGCCACCAGGCGCATAGGCT |
| S105 | TTAATTGCGGGAGAAGAGATACAT |
| S106 | AGACTGTACCCGGAATGATAGCAA |
| S107 | ATTAAAGATTATAAATGATTAGTAATAACATC |
| S108 | TTGGGCGCCAGGGTGGTGATTGCCTCACGACG |
| S109 | TGTAGCCAGCTTTCATCCCCGGTTGGATGTGCCTCGAATTTAATGAAT |
| S110 | TCATTCCATATAACAGTTGATCTAACGGA |
| S111 | TTTTTGCTGAACCTCAAATAATCTAAA |
| S112 | GTAATATCTTGCCCCAAGCTGCAT |
| S113 | GGTGCGGGTGTATAAGGACAGTCAAATCACCA |
| S114 | ATCCCCGGCGGAGAGGATGTCAAT |
| S115 | AGATAACCTGCAGGGATAAAGGAA |
| S116 | GGGCGATCGCACTCCAAATAGGAACGCCATCA |
| S117 | GCTATTAATTAATTTTCCCTTTTCGGTCGAACATGTT |
| S118 | GTTAAATCAGCTCATTAAACGTTAGCGCAACTCACAACATCACATTAA |
| S119 | AAGCTTGCAAGGCTATGAGAATCGATGAACGG |
| S120 | GAGAATAAGAAGCCTTAAATCAAGATTAGTTGCATCGTAGAA TAAGAG |
| S121 | CGACAATGAACCTCCGCAGTATAA |
| S122 | TTTATTATCATTTTATATAATCCTGTTT |
| S123 | AACACTGAAAATATTGACGGAAATTTTACAGA |
| S124 | TTGACAAGATGTTACTATTACGAGTGCAGATACATAACGC |
| S125 | GTAATAAGTTTTAACGCGGAACCGCCTCCCTC |
| S126 | TAGATTTTTTTAATTCGAGCTTCAAAGCGAACCCTCATATATT TTAAATTAAAACAG |
| S127 | CACCCTCACCACCAGAGTTTGCCTTTAGCGTC |
| S128 | CTCAATCAATATCTGGTCAGTTGGAGTGCCAC |
| S129 | CAATCATAAGGGAACCCCAAGTTACAAAATCGATGGTTTGAA ATACCG |
| S130 | ATTCATCAGCGGTTGTTCCAGTTTACGCAAAT |
| S131 | TAAACAACTGAATTTTAACGAGGG |
| S132 | GCCAGGGTAGGTCGACTTTGCGTA |
| S133 | ACACGACCCCTGCAACCAAATCAACAGTTGAAAGGAATTG |
| S134 | AAATCCTGGTCTATCACGCCAGAA |
| S135 | CTAAAAGAGTCAGAGGTCACCCTGTTTCAGCGGAG |
| S136 | ATATCTTTAGGAGCACTAACAACTCGAACGAA |
| S137 | GGCCGGCCACCCTTTGAGGCAAGCAG |
| S138 | AGACTTTTTCATGAGGGAACGCGAGGCGTATTAACTG |
| S139 | CCAATAAAGGCTTGAGATGGTTTAAACGAGTA |
| S140 | AAACTACAGGTTTTGCCAATAGAAGGAACAACGTTAAAGG |
| S141 | TGCCAGAGTGCTTTTTTGGATGGCTTAG |
| S142 | TTTTCTAAAGTTTTGTCGTCTTTCGAAACGGCTACAGAGGCTTT GTTT |
| S143 | AAACTAGCGTAGCTATTTTTGAGAAAGCAATA |
| S144 | TTCTGACCTAAATTTACGCAGAGG |
| S145 | TCAACTAAGCAACGAGAATAAAGATTCATCAG |
| S146 | CCCAATCCAACGCTAACGAGCGTCTTTCCAGATTCCAAGA |
| S147 | ACCGAGGAGCCGAACATAAGACTCTAGCACCATTACCTTT |
| S148 | TCCAATACGCAAATCCTACCACAT |
| S149 | ACTTTTGCTCCTTTTGGAGTAGATTTAGTTTGACCTCAGGATT |
| S150 | GATCTACAATGCCTGCTTTCCCAG |
| S151 | CAATAAACCTGAGGCTCACAAGAA |
| S152 | GCAACAGCTTTTTCTTCCAGTGCC |
| S153 | TACATGGCTTTTGATGGAACCGCCACCCTCAGTGTACCGT |
| S154 | TTTGCACGGCAATGCCTGAGTAAGTGTAA |
| S155 | GAAGCATAAATGTGTATTAAAATT |
| S156 | AACACCCTCAGCCAAAGACAAAAGAGCATTCC |
| S157 | TACCAGTCCCCAAATCAACGTAACGATAAATT |
| S158 | TTCTGACCTGAAAGCGACTTGCCTTATAATCA |
| S159 | CGTAATGGCGTGGGAACAAACGGCTTT |
| S160 | GGGAAGAATGGTCAATCTGCGAACATATAAAT |
| S161 | ACTTTTTCAAATATATAGCAAAAG |
| S162 | ACCAGTACAGAGCCGCAGCGT |
| S163 | GGACAGTAGCGCTGGATTCAGAAATAGTAAGA |
| S164 | GGATTAGGAAAATCACCGGAAACCTATTATTCTGAAA |
| S165 | TATCAGAGAGATATAGTCTGTCCA |
| S166 | TAAATTGTTTTTAACCGCCAGCTTTCCGGCAC |
| S167 | GAGCTGAAGAGCATAAAGCTAAATATTAATGC |
| S168 | TAGCAGCCTATTCATTAACGCAAAGTTGATATA |
| S169 | AGATTTAGGAAAATCGCAAGACAAAAATCGTC |
| S170 | AAAGCGCCGATAGGGTTGAGTGAAGTTATTAA |
| S171 | AATTAGAGCCAGCAAATTATTTATGAAAAGTA |
| S172 | CAACGGAGATTTAACAATTTCATTAATCAATATATGTGAG |
| S173 | TTTAGAAGGAAAAACCAATCAATATTT |
| S174 | TTTAGTATACCGTGTGATAAATAACTGATTGC |
| S175 | GTACCGAGTGCAAGGCAGCGGTCCCCGCCGCGGCCGATTA |
| S176 | TAGCACAATTTCCATTGAATATAAAGTACCGACGAGCCAG |
| S177 | TCGAGAGGGACACCACTTTCACGT |
| S178 | AACAGAGAAAACAGAGTATCTAAA |
| S179 | TTGCTTTGACGAGCACGGATTGACTTGTAAAACGATTT |
| S180 | TTCAAAAGAGGATAAAAATTTTTAGAACCAGA |
| S181 | CGCAATGGAAGGGTTATGATGGCAAGCATCATATTCCTGATTT |
| S182 | ACGGGAGATTTAGCGAACCTCCCGACTTGCGGAGCAAATC |
| S183 | GGTAATTGTGGTTTACCAGCGACGCAAGAGAA |
| S184 | ATTTTCGGTCATAGCCAAGTGCCG |
| S185 | TAGGCAGAGGCATTTTCAAAAGGTGCGGGATC |
| S186 | TCCAAAAAAAAGGCTCGGCAACAT |
| S187 | CATATGTACAACATTATAACCGTG |
| S188 | TAGGTTGGGTTATATAACTATTTCATCAGCCAACG |
| S189 | CGGAATTATCCTGGGGTGCCTAATGAGTGTAGCCCGA |
| S190 | GGAACAAGAGTCCACTCGCTTCTG |
| S191 | GAACAAGCATAGATAAGTCCTGAAATTCTTAC |
| S192 | AGCCCAATGAATCATTCAGCTAAT |
| S193 | TTCGCAAAAAAAGTCAGAAGCAAAGAGTAATC |
| S194 | AGAGAGTACCTTCCGCTCAAGGCCGGACAAATATT |
| S195 | CATTTTTGCTGTTAATTGC |
| S196 | TAGCAATATAATGCT |
| S197 | ATTCGCCAGTGCCGGAAAATTTTT |
| S198 | AAGGGATTTAACACCGAGTAATAAAAGGGACA |
| S199 | TAAGAATAAAAAATACAATAGATTAGAGCCGTCAATAGAT |
| S200 | TTTAAAGAGGCAAAAGAATGCACCAACCTAAAACGTAATGCCA |
| S201 | GCGCAGTCTCTGAATTTACCGCCGCCACCAGAACCATTTTCAGG |
| S202 | AAATATTCGCTTAGAT |

| | | |
|---|---|---|
| | S203 | GCAACACTCGACGCCAGAATC |
| | S204 | TTTCCGGAATCATAATTACTAGAAGCATGTAG |
| | S205 | ACAGACAGGGACCAGAGCCACCACGGGTCAGTGCCTTGAGTAACAGTGCCCGT |
| | S206 | TTTGAATAGAACTGACCAACTTTGAAAGAGTACCTTT |
| | S207 | TTGCGAATAATAATTTGGAATAAGTTGAGTTA |
| | S208 | CGGCCAACGCGCGGGGTTGCAGCA |
| | S209 | TTGTATCGGTTTATCATTAGCAAACTTTTTAA |
| | S210 | TTGCGTTGCGCTCACTAAATCGGC |
| Docking strands | AuL1_12 | AAAAAAAAAAAATTAATTACATTTGTATCATCGCCTAAAGCTGC |
| | AuL2_12 | AAAAAAAAAAAAGTGTCGAAGAAACAAACATCAAGATGTAGAAC |
| | AuL3_12 | AAAAAAAAAAAATCATTCAGGCGATTTTAAGAACTGTGGGGCGC |
| | AuR1_12 | AAAAAAAAAAAACGAATTATACGAGGCGCAGACGGTGGCTGACC |
| | AuR2_12 | AAAAAAAAAAAATTCATCAAGCGGATTGCATCAAAAAAACTCCA |
| | AuR3_12 | AAAAAAAAAAAAGTGTACAGCGAAAGACTTCAAATATCGCGCAGGTTTA |
| | Docking | CAAATGCTTTAAAATCGTCATATAACCCTGGAGGGAGGG |
| Blocking strands | BR1_12 | CGAATTATACGAGGCGCAGACGGTGGCTGACC |
| | BR2_12 | TTCATCAAGCGGATTGCATCAAAAAAACTCCA |
| | BR3_12 | GTGTACAGCGAAAGACTTCAAATATCGCGCAGGTTTA |
| Capping Strands | AuCap_12 | TTTTTTTTTTTT/3ThioMC3-D/ |
| Fluorescent emitter | Cy5 | /5Cy5/CCCTCCCTCC |

Table S3. Effect of Au(III) concentrations on the diameter of synthesized AuNRs

| $AgNO_3$ (μL) | Ascorbic acid (μL) | $HAuCl_4$ (μL) | CTAB solution (μL) | Seed solution (μL) | Measured diameter (nm) |
|---|---|---|---|---|---|
| 50 | 9.4 | 200 | 740.6 | 1.2 | 9.1±0.5 |
| 50 | 12.6 | 267 | 670.4 | 1.2 | 12.1±0.7 |
| 50 | 15.7 | 333 | 601.3 | 1.2 | 14±0.5 |
| 50 | 23.6 | 500 | 426.4 | 1.2 | 17±0.6 |
| 50 | 31.5 | 918.5 | 0 | 1.2 | 19.4±0.3 |

By adjusting the concentration of $HAuCl_4$ while keeping other conditions the same, the width of the AuNRs could be tuned independently. It was found that when Au(III) concentration increased from 0.3 mM to ~0.9 mM, the width of AuNRs raised from 9.1 nm to 19.4 nm.

Table S4. Effect of $AgNO_3$ concentrations on the AR of synthesized AuNRs

| $AgNO_3$ (μL) | Ascorbic acid (μL) | $HAuCl_4$ (μL) | CTAB solution (μL) | Seed solution (μL) | Measured AR |
|---|---|---|---|---|---|
| 40 | 12.6 | 267 | 680.4 | 1.2 | 2.0 |
| 50 | 12.6 | 267 | 670.4 | 1.2 | 2.3 |
| 60 | 12.6 | 267 | 660.4 | 1.2 | 2.5 |
| 70 | 12.6 | 267 | 650.4 | 1.2 | 2.7 |
| 80 | 12.6 | 267 | 640.4 | 1.2 | 2.8 |
| 100 | 12.6 | 267 | 620.4 | 1.2 | 3.0 |

By adjusting the concentration of $AgNO_3$ while keeping other conditions the same, the AR of the AuNRs could be tuned independently. It was found that when Ag concentration increased from 0.04 mM to ~0.1 mM, the AR of AuNRs raised from 2.0 to 3.0.

References

[1] Roller, E. M.; Argyropoulos, C.; Högele, A.; Liedl, T.; Pilo-Pais, M. Plasmon-exciton coupling using DNA templates. *Nano Lett.* **2016**, *16*, 5962-5966.

[2] Puchkova, A.; Vietz, C.; Pibiri, E.; Wünsch, B.; Sanz Paz, M.; Acuna, G. P.; Tinnefeld, P. DNA origami nanoantennas with over 5000-fold fluorescence enhancement and single-molecule detection at 25 μM. *Nano Lett.* **2015**, *15*, 8354-8359.

[3] Han, D.; Pal, S.; Nangreave, J.; Deng, Z.; Liu, Y.; Yan, H. DNA Origami with complex curvatures in three-dimensional space. *Science* **2011**, *332*, 342-346.

[4] Deaton, R.; Garzon, M.; Murphy, R.C.; Rose, J. A.; Franceschetti, D. R.; Stevens Jr, S. E. Reliability and efficiency of a DNA-based computation. *Phys. Rev. Lett.* **1998**, *80*, 417-420.

[5] Samanta, A.; Zhou, Y.; Zou, S.; Yan, H.; Liu, Y. Fluorescence quenching of quantum dots by gold nanoparticles: a potential long range spectroscopic ruler. *Nano Lett.* **2014**, *14*, 5052-5057.

[6] Nangreave, J.; Yan, H.; Liu, Y. DNA Nanostructures as models for evaluating the role of enthalpy and entropy in polyvalent binding. *J. Am. Chem. Soc.* **2011**, *133*, 4490-4497.

[7] Andreani, L. C.; Panzarini, G.; Gérard, J. M. Strong-coupling regime for quantum boxes in pillar microcavities: theory. *Phys. Rev. B* **1999**, *60*, 13276.

[8] Raimond, J. M.; Brune, M.; Haroche, S. Manipulating quantum entanglement with atoms and photons in a cavity. *Rev. Mod. Phys.* **2001**, *73*, 565.

[9] Reithmaier, J. P.; Sęk, G.; Löffler, A.; Hofmann, C.; Kuhn, S.; Reitzenstein, S.; Keldysh, L. V.; Kulakovskii, V. D.; Reinecke, T. L.; Forchel, A. Strong coupling in a single quantum dot-semiconductor microcavity system. *Nature* **2004**, *432*, 197-200.

[10] Hümmer, T.; García-Vidal, F. J.; Martín-Moreno, L.; Zueco, D. Weak and Strong Coupling Regimes in Plasmonic QED. *Physical Review B* **2013**, 87, 115419.

[11] Xu, Y.; Lee, R. K.; Yariv, A. Quantum analysis and the classical analysis of spontaneous emission in a microcavity. *Phys. Rev. A* **2000**, *61*, 033807.

[12] Purcell, E. M. Spontaneous emission probabilities at radio frequencies. In *Confined Electrons and Photons*. Springer: Boston, 1995; pp 839.

[13] Carmichael, H. J.; Brecha, R. J.; Raizen, M. G.; Kimble, H. J.; Rice, P. R. Subnatural linewidth averaging for coupled atomic and cavity-mode oscillators. *Phys. Rev. A* **1989**, *40*, 5516.

[14] Savasta, S.; Saija, R.; Ridolfo, A.; Di Stefano, O.; Denti, P.; Borghese, F. Nanopolaritons: vacuum Rabi splitting with a single quantum dot in the center of a dimer nanoantenna. *ACS Nano* **2010**, *4*, 6369-6376.

[15] Novotny, L. Strong coupling, energy splitting, and level crossings: a classical perspective. *Am. J. Phys.* **2010**, *78*, 1199-1202.

[16] Yoshie, T.; Scherer, A.; Hendrickson, J.; Khitrova, G.; Gibbs, H. M.; Rupper, G.; Ell, C.; Shchekin, O. B.; Deppe, D. G. Vacuum Rabi splitting with a single quantum dot in a photonic crystal nanocavity. *Nature* **2004**, *432*, 200-203.

[17] Claudon, J.; Bleuse, J.; Malik, N. S.; Bazin, M.; Jaffrennou, P.; Gregersen, N.; Sauvan, C.; Lalanne, P.; Gérard, J. M. A highly efficient single-photon source based on a quantum dot in a photonic nanowire. *Nat. Photon.* **2010**, *4*, 174-177.

[18] Wallraff, A.; Schuster, D. I.; Blais, A.; Frunzio, L.; Huang, R. S.; Majer, J.; Kumar, S.; Girvin, S. M.; Schoelkopf, R. J. Strong coupling of a single photon to a superconducting qubit using circuit quantum electrodynamics. *Nature* **2004**, *431*, 162-167.

[19] Törmä, P.; Barnes, W. L. Strong coupling between surface plasmon polaritons and emitters: a review. *Rep. Prog. Phys.* **2014**, *78*, 013901.

[20] Koenderink, A. F. On the use of Purcell factors for plasmon antennas. *Opt. Lett.* **2010**, *35*, 4208-4230.

[21] Maier, S. A. Plasmonic field enhancement and SERS in the effective mode volume picture. *Opt. Express* **2006**, *14*, 1957-1964.

[22] Decker, M.; Staude, I.; Shishkin, I. I.; Samusev, K. B.; Parkinson, P.; Sreenivasan, V. K.A.; Minovich, A., Miroshnichenko, A. E.; Zvyagin, A.; Jagadish, C.; Neshev, D. N.; Kivshar, Y. S. Dual-channel spontaneous emission of quantum dots in magnetic metamaterials. *Nat. Commun.* **2013**, *4*, 2949-2960.

[23] Olmon, R. L.; Slovick, B.; Johnson, T. W.; Shelton, D.; Oh, S-H.; Boreman, G. D.; Raschke, M. B. Optical dielectric function of gold. *Phys. Rev. B* **2012**, *86*, 235147.

[24] Esposito, R.; Altucci, C.; Velotta, R. Analysis of simulated fluorescence intensities decays by a new maximum entropy method algorithm. *J. Fluoresc.* **2013**, *23*, 203-211.

[25] Hoang, T. B.; Akselrod, G. M.; Argyropoulos, C.; Huang, J.; Smith, D. R.; Mikkelsen, M. H. Ultrafast spontaneous emission source using plasmonic nanoantennas. *Nat. Commun.* **2015**, *6*, 7788.

[26] Schlather, A. E.; Large, N.; Urban, A. S.; Nordlander, P.; Halas, N. J. Near-field mediated plexcitonic coupling and giant Rabi splitting in individual metallic dimers. *Nano Lett.* **2013**, *13*, 3281-3286.

[27] Chikkaraddy, R.; de Nijs, B.; Benz, F.; Barrow, S. J.; Scherman, O. A.; Rosta, E.; Demetriadou, A.; Fox, P.; Hess, O.; Baumberg, J. J. Single-molecule strong coupling at room temperature in plasmonic nanocavities. *Nature* **2016**, *535*, 127-130.

[28] Chikkaraddy, R.; Turek, V. A.; Kongsuwan, N.; Benz, F.; Carnegie, C.; van de Goor, T.; de Nijs, B.; Demetriadou, A.; Hess, O.; Keyser, U. F. Baumberg, J. J. Mapping nanoscale hotspots with single-molecule emitters assembled into plasmonic nanocavities using DNA origami. *Nano Lett.* **2017**, *18*, 405-411.